\documentclass[useAMS,usenatbib,usegraphics]{mn2e}
\usepackage{epsf}
\usepackage{epsfig}
\usepackage{subfigure}

\newcommand{\refsim}{\textsc{ref}}
\newcommand{\agn}{\textsc{agn}}
\newcommand{\nocool}{\textsc{nocool}}

\newcommand{\planck}{\textit{Planck}}
\newcommand{\wmap}{\textit{WMAP}}
\newcommand{\xmm}{\textit{XMM-Newton}}
\newcommand{\rexcess}{\textsc{REXCESS}}
\defcitealias{Lin2004a}{Lin et al. 2004}
\defcitealias{Lin2004b}{Lin et al. (2004)}
\defcitealias{Planck2011a}{Planck Early Results}
\defcitealias{Planck2012a}{Planck Intermediate Results}

\title[Simulated cluster populations]{Towards a realistic population of simulated galaxy groups and clusters}

\author[Le Brun, McCarthy, Schaye \& Ponman]{Amandine~M.~C.~Le Brun$^1$\thanks{E-mail: a.m.lebrun@2013.ljmu.ac.uk}, Ian~G.~McCarthy$^1$\thanks{E-mail:
i.g.mccarthy@ljmu.ac.uk}, Joop~Schaye$^{2}$, Trevor~J.~Ponman$^{3}$
\\
$^{1}$Astrophysics Research Institute, Liverpool John Moores University, 146 Brownlow Hill, Liverpool L3 5RF\\
$^{2}$Leiden Observatory, Leiden University, P. O. Box 9513, 2300 RA Leiden, the Netherlands\\
$^{3}$Astrophysics and Space Research Group, School of Physics and Astronomy, University of Birmingham, Edgbaston, Birmingham B15 2TT
}

\begin{document}
\pagerange{\pageref{firstpage}--\pageref{lastpage}} \pubyear{2014}

\maketitle

\label{firstpage}

\begin{abstract}
We present a new suite of large-volume cosmological hydrodynamical simulations called cosmo-OWLS. They form an extension to the OverWhelmingly Large Simulations (OWLS) project, and have been designed to help improve our understanding of cluster astrophysics and non-linear structure formation, which are now the limiting systematic errors when using clusters as cosmological probes. Starting from identical initial conditions in either the \planck~or \wmap7 cosmologies, we systematically vary the most important `sub-grid' physics, including feedback from supernovae and active galactic nuclei (AGN). We compare the properties of the simulated galaxy groups and clusters to a wide range of observational data, such as X-ray luminosity and temperature, gas mass fractions, entropy and density profiles, Sunyaev--Zel'dovich flux, {\it I}-band mass-to-light ratio, dominance of the brightest cluster galaxy, and central massive black hole (BH) masses, by producing synthetic observations and mimicking observational analysis techniques. These comparisons demonstrate that some AGN feedback models can produce a realistic population of galaxy groups and clusters, broadly reproducing both the median trend and, for the first time, the scatter in physical properties over approximately two decades in mass ($10^{13}~\textrm{M}_{\odot} \la M_{500} \la 10^{15}~\textrm{M}_{\odot}$) and 1.5 decades in radius ($0.05 \la r/r_{500} \la 1.5$). However, in other models, the AGN feedback is too violent (even though they reproduce the observed BH scaling relations), implying calibration of the models is required. The production of realistic {\it populations} of simulated groups and clusters, as well as models that bracket the observations, opens the door to the creation of synthetic surveys for assisting the astrophysical and cosmological interpretation of cluster surveys, as well as quantifying the impact of selection effects. 
\end{abstract}

%%%%%%%%%
% Introduction %
%%%%%%%%%

\section{Introduction}

It is widely recognized that galaxy clusters are powerful tools for probing cosmology as well as the non-gravitational physics of galaxy formation (for recent reviews, see~\citealt{Voit2005b,Borgani2011}; \citealt*{Allen2011}; \citealt{Kravtsov2012,Weinberg2013}). The last two decades in particular have witnessed exciting developments in cluster cosmology. The {\it ROSAT} satellite conducted the first all-sky survey of galaxy clusters in X-rays in the early 1990s \citep[RASS;][]{Voges1999} and discovered hundreds of new clusters, both in the nearby and distant Universe. The higher spectral and spatial resolution of {\it Chandra} and \xmm~later led to radical changes in our picture of X-ray clusters (e.g. no evidence for large amounts of cold gas in the central regions, non-isothermal temperature profiles). Simultaneously, large optical cluster catalogues became available from the Sloan Digital Sky Survey \citep[SDSS;][]{York2000} and Sunyaev--Zel'dovich (hereafter SZ) observations progressed from the first reliable detections of individual objects (e.g. \citealt*{Birkinshaw1991}; \citealt{Jones1993,Pointecouteau1999}) to large cosmological surveys with the Atacama Cosmology Telescope \citep{Menanteau2010} and the South Pole Telescope \citep{Vanderlinde2010}, culminating in the first all-sky cluster survey since the RASS, the \planck~survey, whose first results were released in 2011 (\citeauthor{Planck2011a}). The increased size and depth of the surveys allowed for the transition of on-going and up-coming cluster cosmological surveys, such as eRosita \citep{Merloni2012}, Euclid \citep{Laureijs2011} and the Dark Energy Survey \citep{DES2005}, into the `era of precision cosmology', where the systematic errors are now starting to dominate over the statistical ones. The limiting systematic uncertainties now come from our incomplete knowledge of cluster physics, especially of its baryonic aspects, and of non-linear structure formation. Further progress requires the development of increasingly realistic theoretical models and the confrontation of synthetic surveys generated using these models with observational datasets.

The theoretical modelling of the formation and evolution of galaxy groups and clusters has progressed considerably in recent years. For instance, the `cooling catastrophe' (i.e. the general tendency of simulated galaxies and groups and clusters of galaxies to form far too many stars; e.g.\ \citealt{Balogh2001}), which has generally plagued cosmological hydrodynamical simulations since their advent, has largely been overcome in simulations which include feedback from supermassive black holes (e.g. \citealt*{Springel2005b}; \citealt{Sijacki2007,Dubois2010,Fabjan2010,McCarthy2010,McCarthy2011}; \citealt*{Short2013}), while feedback from star formation and supernovae (SNe) is insufficient to halt the development of cooling flows and overly massive central galaxies (e.g. \citealt{Borgani2004}; \citealt*{Nagai2007a}). The observation of X-ray cavities in the intracluster medium (ICM) in the centres of galaxy groups and clusters \citep[for recent reviews, see][]{McNamara2007,Fabian2012} provides strong empirical motivation for the inclusion of AGN in simulations. Recent simulation studies that have implemented AGN feedback have concluded that it also helps to reproduce a number of other important properties of groups and clusters, such as the mean baryon fraction trend with mass (e.g. \citealt*{Bhattacharya2008,Puchwein2008}; \citealt{Fabjan2010,McCarthy2010,Planelles2013}), the mean luminosity--temperature relation (e.g. \citealt{Puchwein2008,Fabjan2010,McCarthy2010,Planelles2014}), and the metallicity and temperature profiles of groups outside of the central regions (e.g. \citealt{Fabjan2010,McCarthy2010,Planelles2014}).

In spite of this progress, no model has yet been able to reproduce the scatter in the global scaling relations over the full range of system total masses from low-mass groups to high-mass clusters, nor the thermodynamic state of the hot gas in the central regions and its scatter. (The latter is another way of saying that models do not reproduce the observed cool core--non-cool core dichotomy.)
This may be signalling that there is still important physics missing from the simulations. In addition, most previous theoretical studies have focused on relatively small samples of clusters using `zoomed' resimulations, rather than trying to simulate large representative populations, and have neglected to factor in important biases (e.g. the effects of gas clumping, deviations from hydrostatic equilibrium, and selection effects) when comparing to the observations, which can affect the qualitative conclusions that are drawn from these comparisons.

The OverWhelmingly Large Simulations project \citep[hereafter OWLS;][]{Schaye2010}, which was a suite of over 50 large cosmological hydrodynamical simulations of periodic boxes with varying `sub-grid' physics, addressed our ignorance of important sub-grid physics and its impact on large representative populations of systems. The main aim of the project was to use simulations to gain insight into the physics of galaxy formation by conducting a systematic study of `sub-grid' physics models and their parameters on representative populations. Using OWLS, \citet{McCarthy2010} showed for the first time that the inclusion of AGN feedback allows the simulations to match simultaneously the properties of the hot plasma and of the stellar populations of local galaxy groups (see also \citealt{Stott2012}). However, due to the finite box size of the OWLS runs (at most $100~h^{-1}$ Mpc on a side), they were not well suited for studying massive clusters, or undertaking a study of the scatter in the observable and physical properties of groups and clusters as a function of mass and redshift. In addition, the original OWLS runs adopted a now out-of-date cosmology (based on the analysis of \wmap~3-year data).

In the present study, we present an extension to the OWLS project (called cosmo-OWLS), consisting of a suite of large-volume cosmological hydrodynamical simulations designed with on-going and up-coming cluster cosmology surveys in mind. The large volumes (here we present simulations in $400~h^{-1}$ Mpc on a side boxes) allow us to extend our comparisons to higher masses and redshifts and to examine the scatter in the physical properties of groups and clusters. The main aims of cosmo-OWLS are: (i) to provide a tool for the astrophysical interpretation of cluster survey data, (ii) to help quantify the group/cluster selection functions that are crucial for cluster cosmology, (iii) to quantify the biases in reconstructed (rather than directly observable) quantities, such as system mass, and the resulting bias in the inferred cosmological parameters, and (iv) to make predictions for future observations. Lastly, we have not only extended the study of \citet{McCarthy2010} to higher masses (and with an updated cosmology), but we have also investigated the effects of baryonic physics upon a larger number of observed properties, such as Sunyaev--Zel'dovich flux, central supermassive black hole scaling relations, and properties of the brightest cluster galaxy of local galaxy groups and clusters.

This paper is organized as follows. We briefly describe the cosmo-OWLS runs, as well as how they were post-processed to produce synthetic observations in Section 2. We then make like-with-like comparisons with global X-ray scaling relations in Section 3.1, and examine the radial distributions of X-ray properties in Section 3.2, followed by an investigation of the Sunyaev--Zel'dovich scalings in Section 4, and of the optical and black hole properties in Section 5. Finally, we discuss and summarize our main findings in Section 6. 

Masses are quoted in physical $\textrm{M}_{\odot}$ throughout.

%%%%%%%%%
% Simulations %
%%%%%%%%%

\section{cosmo-OWLS}
\label{sec:cosmo-owls}

\subsection{Simulation characteristics}
\label{sec:sims}

\begin{table*}
\centering
\begin{tabular}{|l|l|l|l|l|l|l|}
 \hline
	Simulation & UV/X-ray background & Cooling & Star formation & SN feedback & AGN feedback & $\Delta T_{heat}$ \\
	\hline
 \nocool & Yes & No & No & No & No & ...\\
 \refsim & Yes & Yes & Yes & Yes & No & ...\\
 \agn~8.0 & Yes & Yes & Yes & Yes & Yes & $10^{8.0}$ K\\
 \agn~8.5 & Yes & Yes & Yes & Yes & Yes & $10^{8.5}$ K\\
 \agn~8.7 & Yes & Yes & Yes & Yes & Yes & $10^{8.7}$ K\\
 \hline
\end{tabular}
\caption{cosmo-OWLS runs presented here and their included sub-grid physics. Each model has been run in both the \wmap7 and \planck~cosmologies.}
\label{table:owls}
\end{table*}

The original OWLS~runs were limited in size to $100~h^{-1}$ Mpc, with initial conditions based on the 3-year {\it Wilkinson Microwave Anisotropy Probe} (\wmap) maximum-likelihood cosmological parameters \citep{Spergel2007}. The corresponding volume is too small to contain more than a handful of massive clusters of galaxies with $M_{500} \ga 10^{14}~\textrm{M}_\odot$, which have a comoving space density of $\sim 10^{-5}$ Mpc$^{-3}$ at $z=0$ \citep[e.g.][]{Jenkins2001}. With cosmo-OWLS, we are carrying out much larger volume simulations and present here $400 ~h^{-1}$ (comoving) Mpc on a side periodic box simulations with updated initial conditions based either on the maximum-likelihood cosmological parameters derived from the 7-year \wmap~data \citep{Komatsu2011} \{$\Omega_{m}$, $\Omega_{b}$, $\Omega_{\Lambda}$, $\sigma_{8}$, $n_{s}$, $h$\} = \{0.272, 0.0455, 0.728, 0.81, 0.967, 0.704\} or the \planck~data (\citeauthor{Planck_cosmology}) = \{0.3175, 0.0490, 0.6825, 0.834, 0.9624, 0.6711\}. We use the prescription of \citet{Eisenstein1999} to compute the transfer function and the software package {\small N-GenIC}\footnote{http://www.mpa-garching.mpg.de/gadget/} (developed by V.~Springel) based on the Zel'dovich approximation to generate the initial conditions. For each of the models presented below, we have run simulations with both cosmologies. We will only present the results of the \planck~cosmology runs, but comment on any significant differences in the corresponding \wmap7 runs.

The simulations presented here all have $2\times1024^{3}$ particles (as opposed to $2\times512^{3}$ for the original $100~h^{-1}$ Mpc OWLS~volumes), yielding dark matter and (initial) baryon particle masses of $\approx4.44\times10^{9}~h^{-1}~\textrm{M}_{\odot}$ ($\approx3.75\times10^{9}~h^{-1}~\textrm{M}_{\odot}$) and $\approx8.12\times10^{8}~h^{-1}~\textrm{M}_{\odot}$ ($\approx7.54\times10^{8}~h^{-1}~\textrm{M}_{\odot}$), respectively for the \planck~(\wmap7) cosmology. As we have increased the volume by a factor of 64 but `only' increased the number of particles by a factor of 8 with respect to OWLS, the runs presented here are approximately a factor of 8 lower in mass resolution compared to OWLS\footnote{Available hardware prevents us from running higher resolution simulations in such large volumes. A single cosmo-OWLS run has a peak memory consumption of approximately 2.5 TB of RAM, while 6 TB of storage is required for the snapshot data.}. However, as demonstrated in Appendix~\ref{sec:reso} \citep[see also][]{McCarthy2010}, we achieve good convergence in global properties down to halo masses of a few $10^{13}~\textrm{M}_\odot$ at cosmo-OWLS resolution. We note that the gravitational softening of the runs presented here is fixed to $4~h^{-1}$ kpc (in physical coordinates below $z=3$ and in comoving coordinates at higher redshifts).

As the hydrodynamic code and its sub-grid physics prescriptions used for cosmo-OWLS have not been modified from that used for OWLS, and have been described in detail in previous papers, we present only a brief description below.

The simulations were carried out with a version of the Lagrangian TreePM-SPH code \textsc{gadget3} \citep[last described in][]{Springel2005a}, which has been significantly modified to include new `sub-grid' physics. Radiative cooling rates are computed element by element, using the method of \citet*{Wiersma2009a}, by interpolating as a function of density, temperature and redshift from pre-computed tables, that were generated with the publicly available photoionization package \textsc{cloudy} \citep[last described in][]{Ferland1998} and calculated in the presence of the CMB and of the \citet{Haardt2001} ultra-violet (UV) and X-ray photoionizing backgrounds. Reionization is modelled by switching on the UV background at $z=9$. Star formation (SF) is implemented stochastically following the prescription of \citet{Schaye2008}. Since the simulations lack both the physics and the resolution to model the cold interstellar medium (ISM), an effective equation of state (EOS) is imposed with $P\propto \rho^{4/3}$ for gas with $n_{H}>n_{H}^{*}$ where $n_{H}^{*}=0.1~\textrm{cm}^{-3}$, and only gas on the effective EOS is allowed to form stars, at a pressure-dependent rate which reproduces the observed Kennicutt-Schmidt SF law without requiring any tuning \citep[see][]{Schaye2008}. Stellar evolution and chemical enrichment are implemented using the model of \citet{Wiersma2009b}, which computes the timed-release of 11 elements (H, He, C, N, O, Ne, Mg, Si, S, Ca and Fe, which represent all of the important ones for radiative cooling) due to both Type Ia and Type II supernovae (SNe) and Asymptotic Giant Branch stars.

Feedback from SNe is implemented using the local kinetic wind model of \citet{DallaVecchia2008} with the initial mass-loading factor and the initial wind velocity chosen to be respectively $\eta=2$~and~$v_{w}=600~\textrm{km s}^{-1}$. These parameter values correspond to a total wind energy which is approximately 40 per cent of the total energy available for the \citet{Chabrier2003} initial mass function (IMF) used by the simulations. Note that the hot gas properties of galaxy groups and clusters are generally insensitive to these parameters, since SN feedback is ineffective at these high masses (i.e. the entropy SNe inject is small compared to that generated by gravitational shock heating or removed by radiative losses).

Three of the runs we present here include AGN feedback due to accretion of matter on to supermassive black holes (BHs). This is incorporated using the sub-grid prescription of \citet{Booth2009}, which is a modified version of the model of \citet{Springel2005b}. The main features of this model are summarized below. 

During the simulation, an on-the-fly friends-of-friends (FoF) algorithm is run on the dark matter distribution. New haloes with more than 100 particles (corresponding to a mass of $\log_{10}[M_{FoF}(\textrm{M}_\odot/h)] \approx 11.6$) are seeded with black hole sink particles with an initial mass that is 0.001 times the gas particle mass. Note that this is the same prescription as used for the OWLS AGN model (see \citealt{Booth2009}). The fixed dark matter particle number for seeding implies that BHs are injected into more massive haloes (by approximately a factor of 8) in cosmo-OWLS compared to OWLS. In Appendix~\ref{sec:reso}, we compare the growth histories of black hole particles using the OWLS and cosmo-OWLS BH seeding schemes. 

BHs can grow via (Eddington-limited) Bondi--Hoyle--Lyttleton accretion and through mergers with other BHs. Since the simulations lack the physics and resolution to model the cold ISM, they will generally underestimate the true Bondi accretion rate on to the BH by a large factor. Recognizing this issue, \citet{Springel2005b} (and most studies which have adopted this model since then) scaled the Bondi rate up by a constant factor $\alpha \sim 100$. The \citet{Booth2009} model which we adopt, however, has $\alpha$ vary as a power-law of the local density for gas above the SF threshold $n_{H}^{*}$. The power-law exponent $\beta$ is set to 2 and the power-law is normalized so that $\alpha=1$ for densities equal to the SF threshold. Thus, at low densities, which {\it can} be resolved and where no cold interstellar phase is expected, the accretion rate asymptotes to the true Bondi rate.

A fraction of the rest-mass energy of the gas accreted on to the BH is used to heat neighbouring gas particles, by increasing their temperature. As discussed in detail by \citet{DallaVecchia2008,DallaVecchia2012}, thermal feedback in cosmological simulations, be it from SNe or BHs, has traditionally been inefficient: as the feedback energy is being injected into a large amount of mass, it can only raise the temperature of the gas by a small amount. The feedback energy is then radiated away quickly because of the short post-heating cooling time. In nature, the energy is injected into a much smaller mass of gas and thus the post-heating cooling time is typically very long. While much algorithmic progress has been made recently to overcome this problem in the context of SN feedback (see \citealt{DallaVecchia2012} for discussion), less attention has been devoted to this artificial overcooling problem in the context of AGN feedback. The \citet{Booth2009} model overcomes this problem by increasing the temperature of the gas by a pre-defined level $\Delta T_{heat}$. More specifically, a fraction $\epsilon$ of the accreted energy heats up a certain number $n_{heat}$ of randomly chosen surrounding gas particles (within the SPH kernel which contains 48 particles) by increasing their temperature by $\Delta T_{heat}$, with the BHs {\it storing} the feedback energy until it is large enough to heat the $n_{heat}$ particles by $\Delta T_{heat}$. These two parameters are chosen such that the heated gas has a sufficiently long cooling time and that the time needed to have a feedback event is shorter than the Salpeter time for Eddington-limited accretion. \citet{Booth2009} found that $\Delta T_{heat}=10^{8}~\textrm{K}$ and $n_{heat}=1$ correspond to a good balance between these two constraints. These values were hence used for the OWLS `AGN' model. This model is hereafter referred to as \agn~8.0.

The efficiency $\epsilon$ is set to 0.015, which results in a good match to the normalization of the $z=0$ relations between BH mass and stellar mass and velocity dispersion (the slopes of the relations are largely independent of $\epsilon$), as well as to the observed cosmic BH density, as demonstrated by \citet{Booth2009,Booth2010} (see also Appendix~\ref{sec:reso}). \citet{McCarthy2011} found that galaxy groups simulated using this model for AGN feedback are fairly insensitive to the choice of $\beta$ and $n_{heat}$, whilst they are sensitive to $\Delta T_{heat}$, particularly if it is similar to, or smaller than the group's virial temperature. In the latter cases, the feedback will be inefficient. It is worth noting that the most massive systems expected in the much larger simulated volumes presented here will have $\Delta T_{heat}\sim T_{vir}$. AGN feedback is therefore anticipated to become less efficient for these systems. This has led us to try two additional runs, with increased heating temperatures (leaving $n_{heat}$ and $\epsilon$ fixed): $\Delta T_{heat}=3\times10^{8}~\textrm{K}$ (hereafter \agn~8.5) and $\Delta T_{heat}=5\times10^{8}~\textrm{K}$ (hereafter \agn~8.7). Note that since the same amount of gas is being heated in these models as in the \agn~8.0 model, more time is required for the BHs to accrete enough mass to be able to heat neighbouring gas to a higher temperature. Thus, increasing the heating temperature leads to more bursty and more energetic feedback events.

Table~\ref{table:owls} provides a list of the new runs presented here and the sub-grid physics that they include.

\subsection{Post-processing}

\subsubsection{Halo properties}

Haloes are identified by using a standard friends-of-friends percolation algorithm on the dark matter particles with a typical value of the linking length in units of the mean interparticle separation (b=0.2). The baryonic content of the haloes is identified by locating the nearest DM particle to each baryonic (i.e. gas or star) particle and associating it with the FoF group of the DM particle. Artificial haloes are removed by performing an unbinding calculation with the \textsc{subfind} algorithm \citep{Springel2001,Dolag2009}: any FoF halo that does not have at least one self-bound substructure (called subhalo) is removed from the FoF groups list. A `galaxy' is a collection of star and gas particles bound to a subhalo. A halo can thus host several galaxies.

Spherical overdensity masses $M_{\Delta}$ (where $M_{\Delta}$ is the total mass within a radius $r_{\Delta}$ that encloses a mean internal overdensity of $\Delta$ times the critical density of the Universe) with $\Delta=200$, 500 and 2500 have been computed (total, gas and stars) for all the FoF haloes. The spheres are centred on the position of the most bound particle of the main subhalo (the most massive subhalo of the FoF halo). Then, all galaxy groups and clusters with $M_{500}\ge10^{13}~\textrm{M}_{\odot}$ are extracted from each snapshot for analysis. There are roughly $14,000$ such systems at $z=0$ in the \nocool~run with the \planck~cosmology, for example.

\subsubsection{X-ray observables and analysis}
\label{sec:syntheticxray}

It has been demonstrated in a number of previous studies that there can be non-negligible biases in the derived hot gas properties (e.g. due to multi-temperature structure and clumping) and system mass (e.g. $M_{500}$) inferred from X-ray analyses (e.g. \citealt{Mathiesen2001,Mazzotta2004,Rasia2006}; \citealt*{Nagai2007b}; \citealt{Khedekar2013}). Thus, to make like-with-like comparisons with X-ray observations, we produce synthetic X-ray data and then analyse them in a way that is faithful to what is done for the real data. Below we describe our procedure for producing and analysing synthetic X-ray observations.

For each hot gas particle within $r_{500}$, we compute the X-ray spectrum in the 0.5--10.0 keV band using the Astrophysical Plasma Emission Code \citep[APEC;][]{Smith2001} with updated atomic data and calculations from the AtomDB v2.0.2 \citep{Foster2012}. The spectrum of each gas particle is computed using the particle's density, temperature, and full abundance information. More specifically, for each particle, we compute a spectrum for each of the 11 elements tracked by the simulations, we scale each spectrum appropriately using the particle's elemental abundances (the fiducial APEC spectrum assumes the Solar abundances of \citealt{Anders1989}), and we sum the individual element spectra to create a total spectrum for the particle. Note that we exclude cold gas below $10^5$ K which contributes negligibly to the total X-ray emission. We also exclude any (hot or cold) gas which is bound to self-gravitating substructures (`subhaloes'), as observers also typically excise substructures from their X-ray data. Note that the smallest subhaloes that can be resolved in the present simulations have masses $\sim 10^{11}~\textrm{M}_\odot$.

Gas density, temperature, and metallicity profiles are `measured' for each simulated system by fitting single-temperature APEC models with a metallicity that is a fixed fraction of Solar (as commonly assumed in observational studies) to spatially-resolved X-ray spectra in (three-dimensional) radial bins. (Note that the observed radial profiles we compare to in Section 3.2 are all derived under the assumption of spherical symmetry.) The radial bins are spaced logarithmically and we use between 10-20 bins within $r_{500}$ (similar to what is possible for relatively deep {\it Chandra} observations of nearby systems). To more closely mimic the actual data quality and analysis, the cluster spectra (and the single-temperature APEC model spectra to be fitted to the cluster spectra) are multiplied by the effective area energy curve of {\it Chandra}, subjected to Galactic absorption due to HI with a typical column density of $2\times10^{20}$ cm$^2$, and re-binned to an energy resolution of 150 eV (i.e. similar to the {\it Chandra} energy resolution). The single-temperature model spectra are fitted to the cluster spectra using the \textsc{mpfit} least-squares package in \textsc{idl} \citep{Markwardt2009}. In general, including a Galactic absorption column and multiplying by the effective energy curve of {\it Chandra} have only a very small effect (a few per cent) on the recovered density, temperature and metallicity profiles, by affecting which parts of the spectra are most heavily weighted in the fit. 

In addition to profiles, we also derive `mean' system X-ray temperatures and metallicities by following the above procedure but using only a single radial bin: either [0--1]$r_{500}$ (`uncorrected') or [0.15--1]$r_{500}$ (`cooling flow-corrected'). System X-ray luminosities within $r_{500}$ are computed in the soft $0.5-2.0$ keV band by summing the luminosities of the individual particles within this three-dimensional radius (the luminosity of an individual particle is computed by integrating the particle's spectrum over this band).

When making comparisons to X-ray-derived mass measurements (e.g. $M_{500}$), we employ a hydrostatic mass analysis of our simulated systems using the measured gas density and temperature profiles inferred from our synthetic X-ray analysis described above. In particular, we fit the density and temperature profiles using the functional forms proposed by \citet{Vikhlinin2006} and assume hydrostatic equilibrium to derive the mass profile. We will use the subscript `hse' to denote quantities inferred from (virtual) observations under the assumption of hydrostatic equilibrium. 

In Appendix~\ref{sec:biases}, we explore the sensitivity of the HSE and spectroscopic temperature biases (and scatter about the bias) to sub-grid physics but defer a detailed analysis of these biases to a future study.

\subsubsection{`Optical' observables}
\label{sec:OptObs}

Optical and near-infrared luminosities and colours are computed using the \textsc{galaxev} model of \citet{Bruzual2003} to derive a spectral energy distribution for each star particle, which is then convolved with the transmission function of the chosen band filter. When doing so, each star particle is treated as a simple stellar population with a \citet{Chabrier2003} IMF and the star particle's age and metallicity. We ignore the effects of dust attenuation but compare to dust-corrected observations where possible.

%%%%%%%%%%%
% X-ray properties %
%%%%%%%%%%%

\section{X-ray properties}
\label{sec:Xrayprops}

We begin by comparing the X-ray properties of the simulated groups and clusters to observations of local ($z \sim 0$) systems. In Section~\ref{sec:Xrayscalings}, we examine global hot gas properties as a function of system mass and, in Section~\ref{sec:prof}, we compare to the observed radial distributions of entropy and density.

For clarity, we have omitted observational error bars from the global hot gas property plots below. For reference, the typical {\it statistical} errors are of order 10 per cent in gas mass and temperature, 5 per cent in X-ray luminosity, and 10--20 per cent in halo mass for the observational samples we compare to below. For the same reason, we have also only plotted the scatter (using shaded regions) for the \agn~8.0 model as the intrinsic scatter does not vary much between the different physical models. 

%% Global scalings

\subsection{Global scaling relations}
\label{sec:Xrayscalings}

\subsubsection{Luminosity--mass relation}
\label{sec:mass_Lx}

\begin{figure*}
\begin{center}
\includegraphics[width=0.49\hsize]{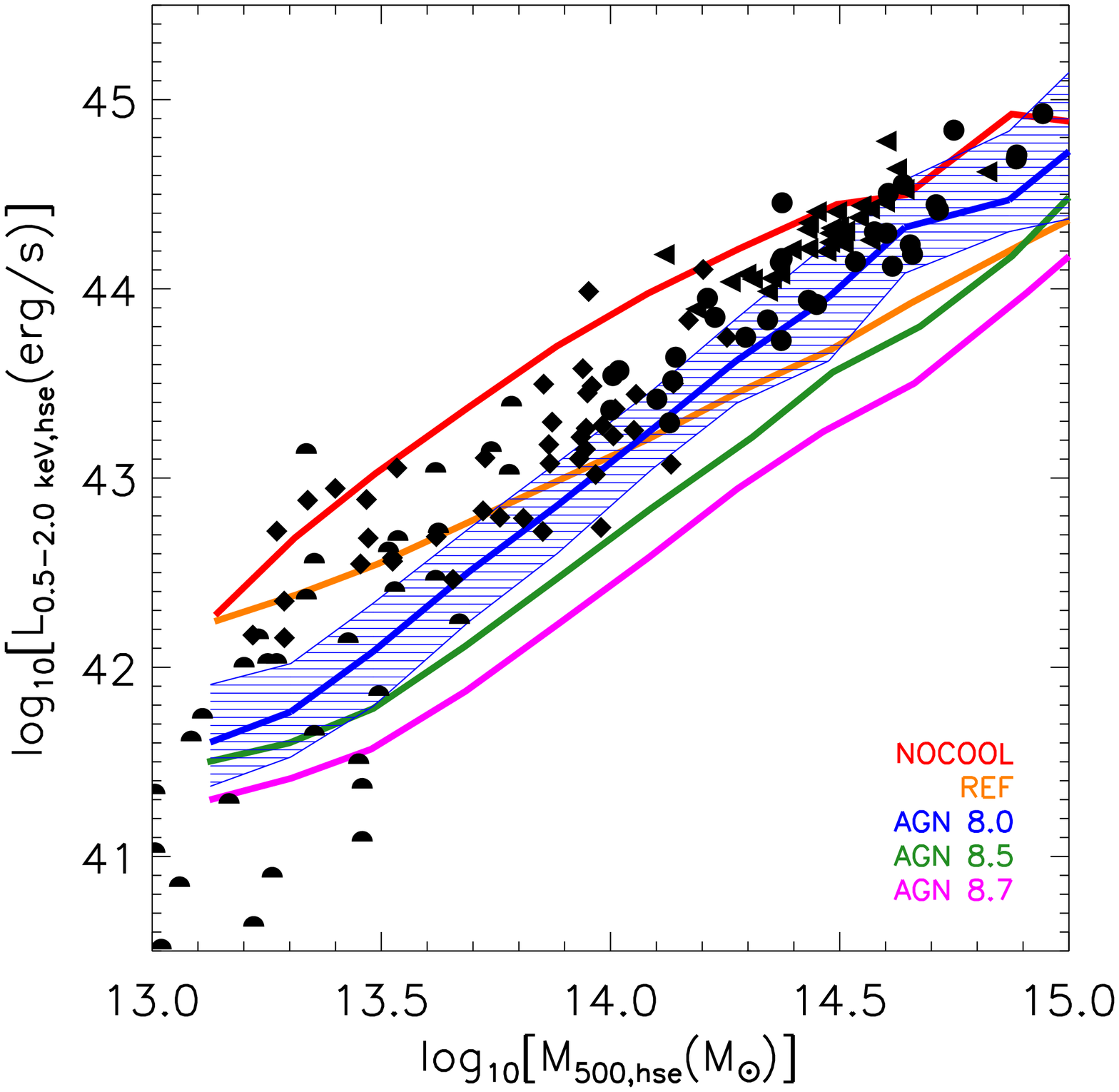}
\includegraphics[width=0.49\hsize]{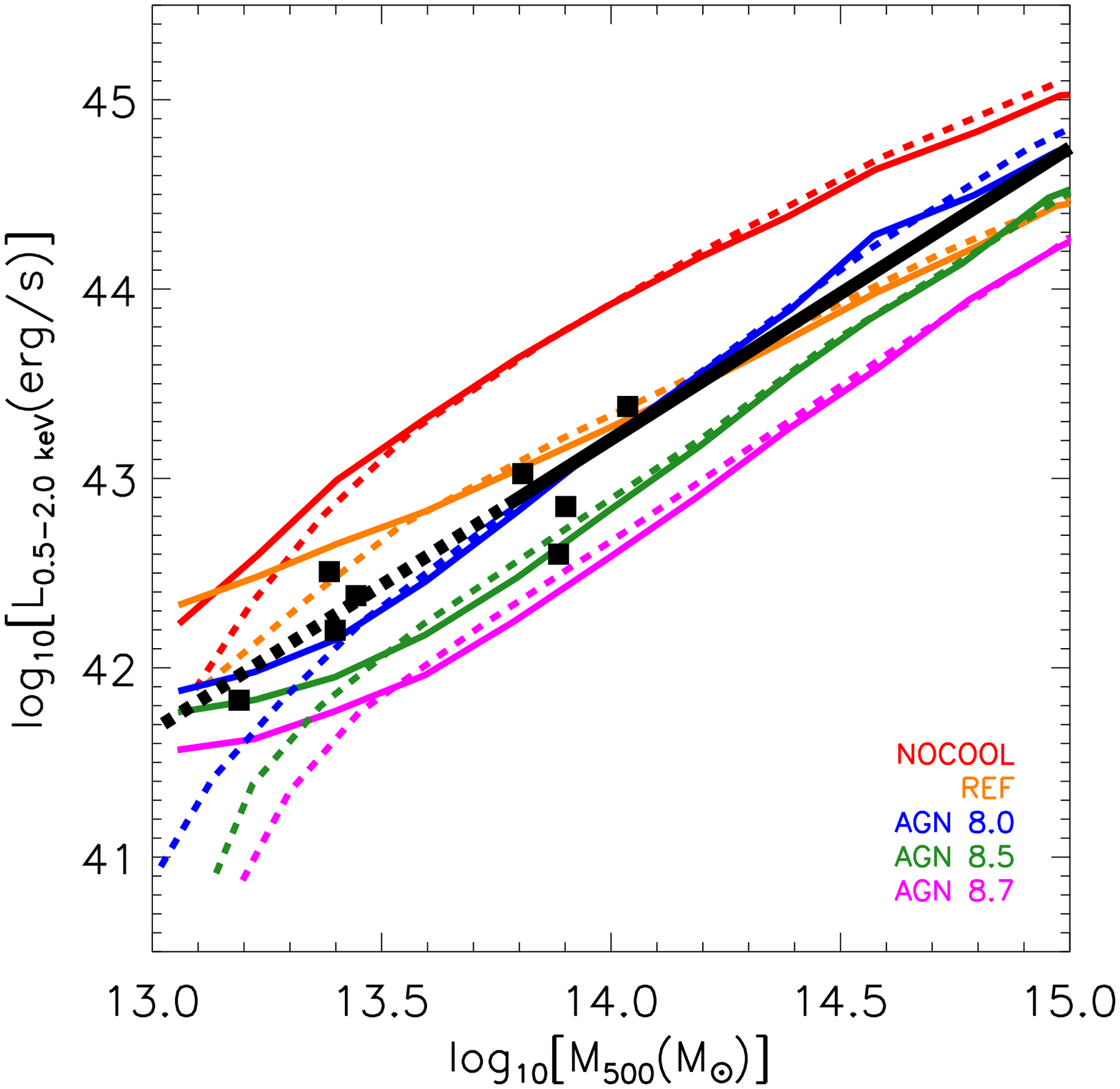}
\caption{The soft X-ray luminosity--$M_{500}$ relation at $z=0$. The X-ray luminosity refers to the 0.5--2.0 keV band (rest-frame) and is computed respectively within $r_{500,hse}$ for the left panel and within $r_{500}$ for the right one. {\it Left:} The filled black circles (clusters), left-facing triangles (clusters), diamonds (groups), and semi-circles (groups) represent the observational data (at $z\approx0$) of \citet{Pratt2009}, \citet{Vikhlinin2009}, \citet{Sun2009} and \citet{Osmond2004}, respectively. The solid curves (red, orange, blue, green and magenta) represent the median $L_X-M_{500,hse}$ relations in bins of $M_{500,hse}$ for the different simulations at $z=0$ and the blue shaded region encloses 68 per cent of the simulated systems for the \agn~8.0 model. {\it Right:} The solid and dashed black lines represent the stacked relation and its extrapolation down to lower masses of \citet{Rozo2009} at $z\approx0.25$, derived by stacking X-ray (RASS) and weak lensing (SDSS) data in bins of richness for the optically-selected maxBCG sample. The filled black squares represent the stacked relation of \citet{Leauthaud2010} scaled to $z=0.25$, which uses stacked weak lensing masses for COSMOS groups in bins of X-ray luminosity for a sample of X-ray-selected groups. The solid and dashed curves (red, orange, blue, green and magenta) represent the simulated mean X-ray-luminosity$-M_{500}$ relations at $z=0.25$ in bins of $M_{500}$ and $L_X$, respectively. The AGN model with a heating temperature of $\Delta T_{heat}=10^{8}$ K (i.e. \agn~8.0) reproduces the observed relations relatively well. Higher heating temperatures (i.e. more bursty feedback) lead to under-luminous systems, while lack of AGN feedback altogether (\refsim) results in over-luminous groups and under-luminous clusters.}
\label{fig:mass_Lx} 
\end{center}
\end{figure*}

In Fig.~\ref{fig:mass_Lx}, we plot the soft (0.5--2.0 keV band) X-ray luminosity--$M_{500}$ relation for the various simulations (coloured solid curves and shaded region) and compare to observations of individual X-ray-selected systems (data points in left panel) and stacking measurements of the mass--luminosity relation for the optically-selected maxBCG sample (\citealt{Rozo2009}; black lines in right panel) and the X-ray-selected COSMOS sample (\citealt{Leauthaud2010}; data points in right panel). As the observational mass measurements of the data in the left panel of Fig.~\ref{fig:mass_Lx} are based on a hydrostatic analysis of the X-ray observations, we use our synthetic X-ray observation methodology described in Section~\ref{sec:syntheticxray} to measure $M_{500,hse}$ for the simulated systems. The maxBCG and COSMOS data in the right panel, on the other hand, use stacked weak lensing masses (in bins of richness and X-ray luminosity, respectively). We use the true $M_{500}$ for the simulated systems in this comparison, as weak lensing masses are thought to be biased on average by only a few per cent (e.g. \citealt*{Becker2011,Bahe2012}, but see \citealt{Rasia2012} who find somewhat larger biases).
 
For the \citet{Leauthaud2010} data, we have converted their 0.1--2.4 keV luminosities into 0.5--2.0 keV luminosities using the online WebPIMMS\footnote{http://heasarc.gsfc.nasa.gov/Tools/w3pimms\_pro.html} tool (the conversion factor is $\approx 0.6$ and is insensitive to the temperature adopted for the range considered here). We have converted their $M_{200}$ masses into $M_{500}$ assuming a NFW profile with a concentration of $4$ \citep[e.g.][]{Duffy2008}, which yields $M_{500} \approx 0.69 M_{200}$. Finally, we have scaled their luminosities and masses to $z=0.25$ assuming self-similar evolution (many of the COSMOS groups are close to this redshift in any case), to be directly comparable to the \citet{Rozo2009} relation and the simulations presented in the right panel of Fig.~\ref{fig:mass_Lx}.

\begin{figure*}
\begin{center}
\includegraphics[width=0.49\hsize]{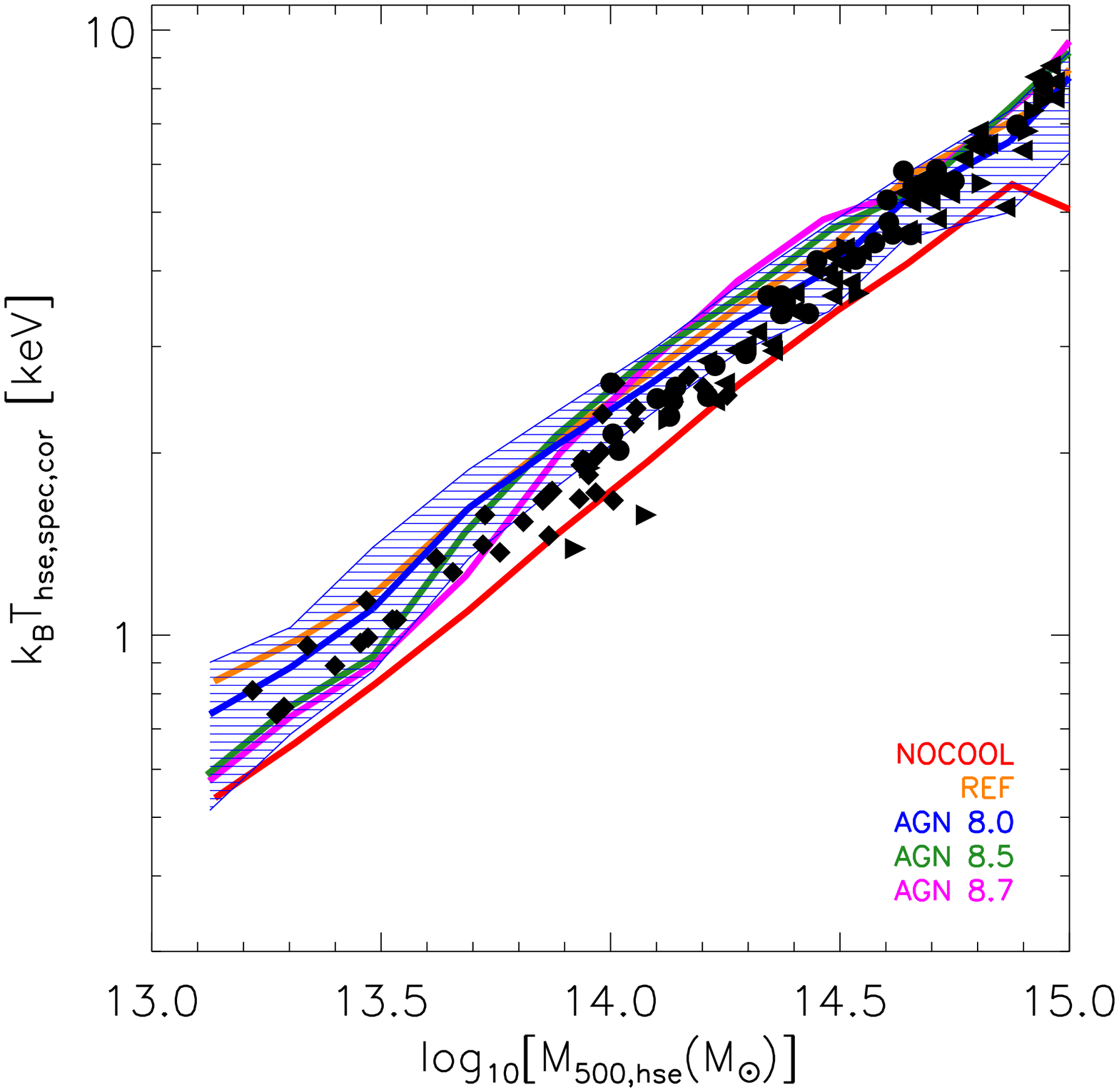}
\includegraphics[width=0.49\hsize]{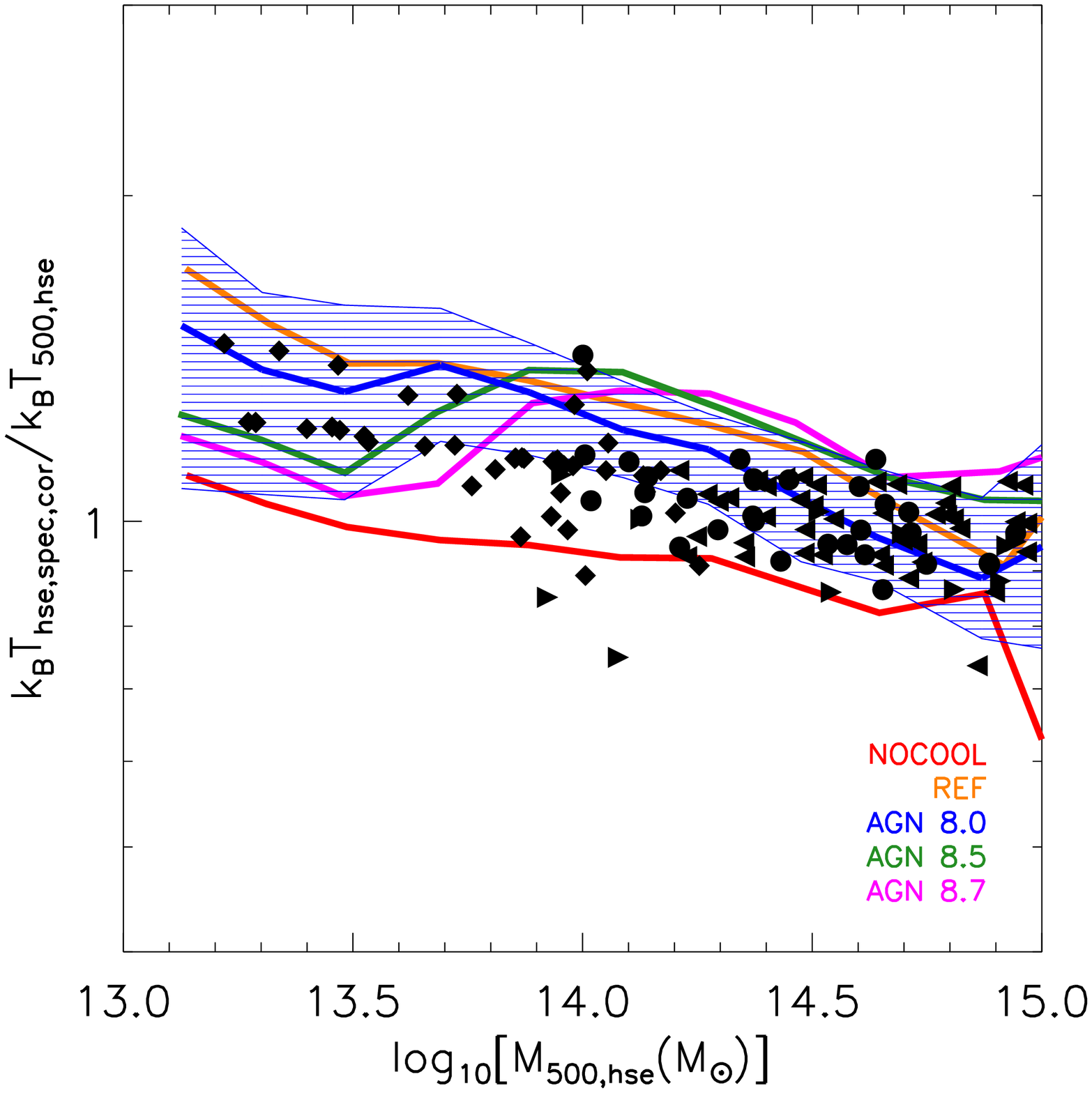}
\caption{The X-ray temperature--$M_{500,hse}$ relation at $z=0$. The X-ray temperature is measured by fitting a single-temperature plasma model to the X-ray spectrum within the annulus [0.15--1]$r_{500,hse}$ (i.e. a mean `cooling flow-corrected' temperature). The filled black circles (clusters), right-facing triangles (clusters), left-facing triangles (clusters), and diamonds (groups) represent the observational data of \citet{Pratt2009}, \citet{Vikhlinin2009}, \citet{Vikhlinin2006} and \citet{Sun2009}, respectively. The coloured solid curves represent the median mass--temperature relations in bins of $M_{500,hse}$ for the different simulations and the blue shaded region encloses 68 per cent of the simulated systems for the \agn~8.0 model. In the left panel, we plot the observed temperature (in keV), while in the right panel the temperature is normalized by the virial temperature $k_BT_{500,hse} \equiv \mu m_p G M_{500,hse} / 2 r_{500,hse}$ to take out the gravitational halo mass dependence. The \agn~8.0 and \refsim~models broadly reproduce the observed relations, while the non-radiative (\nocool) and AGN models with higher heating temperatures (\agn~8.7 in particular) under- and overshoot (respectively) the observed relation by about 10 per cent.} 
\label{fig:mass_Tx}
\end{center}
\end{figure*}

The AGN feedback model with the `standard' OWLS heating temperature of $\Delta T_{heat}=10^{8}$ K (i.e. \agn~8.0, or just `AGN' in \citealt{McCarthy2010}) broadly reproduces the observed luminosity-mass relation over nearly two orders of magnitude in mass. There is a slight difference in slope with respect to the individual X-ray-selected systems in the left panel of Fig.~\ref{fig:mass_Lx}, such that the lowest mass observed systems are a factor of a few more luminous than their simulated counterparts. However, no such offset is evident in the comparison to the stacking results in the right panel, which suggests that observational selection may be important (see discussion below). Interestingly, when we examine the same model in the \wmap7 cosmology, we find that the discrepancy in the left panel largely goes away (the simulated clusters are brighter, presumably due to the increased baryon fraction in the \wmap7 cosmology), although one is introduced in the right panel, in the sense that the simulated clusters become slightly brighter on average than the maxBCG/COSMOS stacking results indicate. 

Increasing the AGN heating temperature significantly (e.g. \agn~8.7, magenta solid line), which makes the AGN feedback more violent and bursty in nature, tends to result in under-luminous systems at all mass scales, independent of our choice of cosmology. As we show in Section~\ref{sec:rhoprof}, this lower luminosity is due to a strong reduction in the central gas density. Neglect of AGN feedback altogether (\refsim) results in a flatter than observed luminosity--mass relation, such that groups (clusters) are over-luminous (under-luminous) with respect to the observations. 

Previous simulation studies, such as those of \citet{Puchwein2008}, \citet{Fabjan2010}, \citet{Short2013} and \citet{Planelles2014} have also concluded that the inclusion of AGN feedback helps to reproduce the mean luminosity--temperature relation.

Interestingly, the observed scatter in the luminosity-mass relation is also broadly reproduced by the models from $\log_{10}[M_{500}(\textrm{M}_\odot)] \ga 14$ or so. This suggests that the simulations have produced reasonably realistic {\it populations} of clusters. At lower masses ($\log_{10}[M_{500}(\textrm{M}_\odot)] \la 13.5$), the observed scatter appears to be considerably larger than in the \agn~8.0 model. This could indicate either the impact of selection effects (see discussion below) in observed surveys, or that the history of AGN activity is more variable in low-mass systems than is allowed by the models. In a future study, we plan to perform a more careful comparison of the observed and simulated scatter and to determine the origin of the scatter in the simulated population.

Note that while we have attempted to `measure' the X-ray properties of our simulated systems in an observational manner, an important caveat to bear in mind is that we have not attempted to {\it select} our systems in the same way as in the observational samples (which generally have poorly understood selection functions). This may affect the quantitative conclusions that can be drawn from comparisons of X-ray luminosities, particularly for galaxy groups where observations of individual groups (left panel) are typically limited to the very brightest and nearest systems \citep{Rasmussen2006}. 

Indeed, there appears to be a noticeable difference in the mean X-ray luminosity of groups (with masses $M_{500}\sim10^{13-13.5}~\textrm{M}_{\odot}$) for the different observational studies. In particular, the \citet{Sun2009} X-ray-selected sample (black diamonds in the left panel) has a significantly higher mean luminosity than the \citet{Osmond2004} X-ray-selected sample (black semi-circles in the left panel), the \citet{Rozo2009} optically-selected sample (black lines in the right panel), and the \citet{Leauthaud2010} X-ray-selected sample (black squares in the right panel). The \citet{Sun2009} sample is based on archival data with the requirement that there be a sufficiently large number of photons to measure spatially-resolved spectra (and therefore temperature and density profiles) out to a significant fraction of $r_{500}$. The \citet{Osmond2004} study, on the other hand, required only enough photons to measure a single mean temperature, which we have converted to $M_{500}$ using the mass--temperature relation of \citet{Sun2009}. For the \citet{Leauthaud2010} sample, galaxy groups only need be detected and have a robust X-ray luminosity (i.e. they do not require a temperature measurement) to be considered in their stacking analyses. Finally, the \citet{Rozo2009} sample is optically-selected and mean X-ray luminosities are derived by stacking shallow RASS X-ray data of many groups and clusters (contamination due to point sources and to false groups may be an issue at such low richnesses, however).

In the future, large samples of homogeneously analysed and selected X-ray groups will be available through the XXL \citep{Pierre2011} and eRosita \citep{Merloni2012} surveys. Particular attention is being devoted in these surveys to the selection function using synthetic observations of cosmological simulations. For the present, the importance of selection remains an open question for the observed mass--luminosity and luminosity--temperature relations.

\subsubsection{Mass--temperature relation}

In the left panel of Fig.~\ref{fig:mass_Tx}, we plot the $M_{500,hse}-$X-ray temperature relation at $z=0$ for the various simulations and compare to observations of individual X-ray-selected systems. For both the observations and simulations, the X-ray temperature is measured by fitting a single-temperature plasma model to the integrated X-ray spectrum within the annulus [0.15--1]$r_{500,hse}$ (i.e. a mean `cooling flow-corrected' temperature). In the right panel of Fig.~\ref{fig:mass_Tx}, the temperature has been normalized by the virial temperature $k_BT_{500,hse}\equiv\mu m_p G M_{500,hse} / 2 r_{500,hse}$ to take out the explicit gravitational halo mass dependence, in order to more closely examine the effects of baryonic physics on the mass--temperature relation. Note that the virial temperature is computed using the hydrostatically-derived mass for both the observed and simulated systems.

The mass--temperature relation is similar for all the runs we have examined and independent of the choice of cosmology. This insensitivity owes to the fact that, to first order, the temperature is set by the depth of the potential well, which is dominated by dark matter. As a result, the X-ray temperature is always close to the virial temperature (as demonstrated in the right panel), particularly for the core-excised temperatures\footnote{Coalescence of baryons can potentially become gravitationally important in the very central regions of simulations that suffer from overcooling (e.g. \citealt{Nagai2007a} and the \refsim~model here), which can lead to strong gravitational compression and `heating'.} used in Fig.~\ref{fig:mass_Tx}, which probe gas with long cooling times. This is consistent with the findings of previous simulation studies (e.g. \citealt{McCarthy2010,Short2010}).

The \nocool~model lies below the observed relation by roughly 10 per cent (i.e. it has too low temperatures at fixed masses compared to the observations). As we will show below, this is because the ICM has a lower entropy in this run compared to the other runs, due both to its inability to cool (which would remove the lowest-entropy gas from the ICM) and to the lack of feedback (which heats and ejects low-entropy gas). On the other hand, AGN models with high heating temperatures (\agn~8.5 and, in particular, \agn~8.7) lie above the observed relation at $z=0$ because they eject too much low-entropy gas. Finally, there is slight difference in the shape of the relations predicted by all the radiative simulations compared to the observations, with a `bump' in the median trends of the simulations at $\log_{10}[M_{500}(\textrm{M}_\odot)]\sim14$. This is due to the differences in detailed entropy structure of the gas between the simulations and observations (see Figs.~\ref{fig:Sprof} and \ref{fig:Sref}).

\begin{figure}
\begin{center}
\includegraphics[width=1.0\hsize]{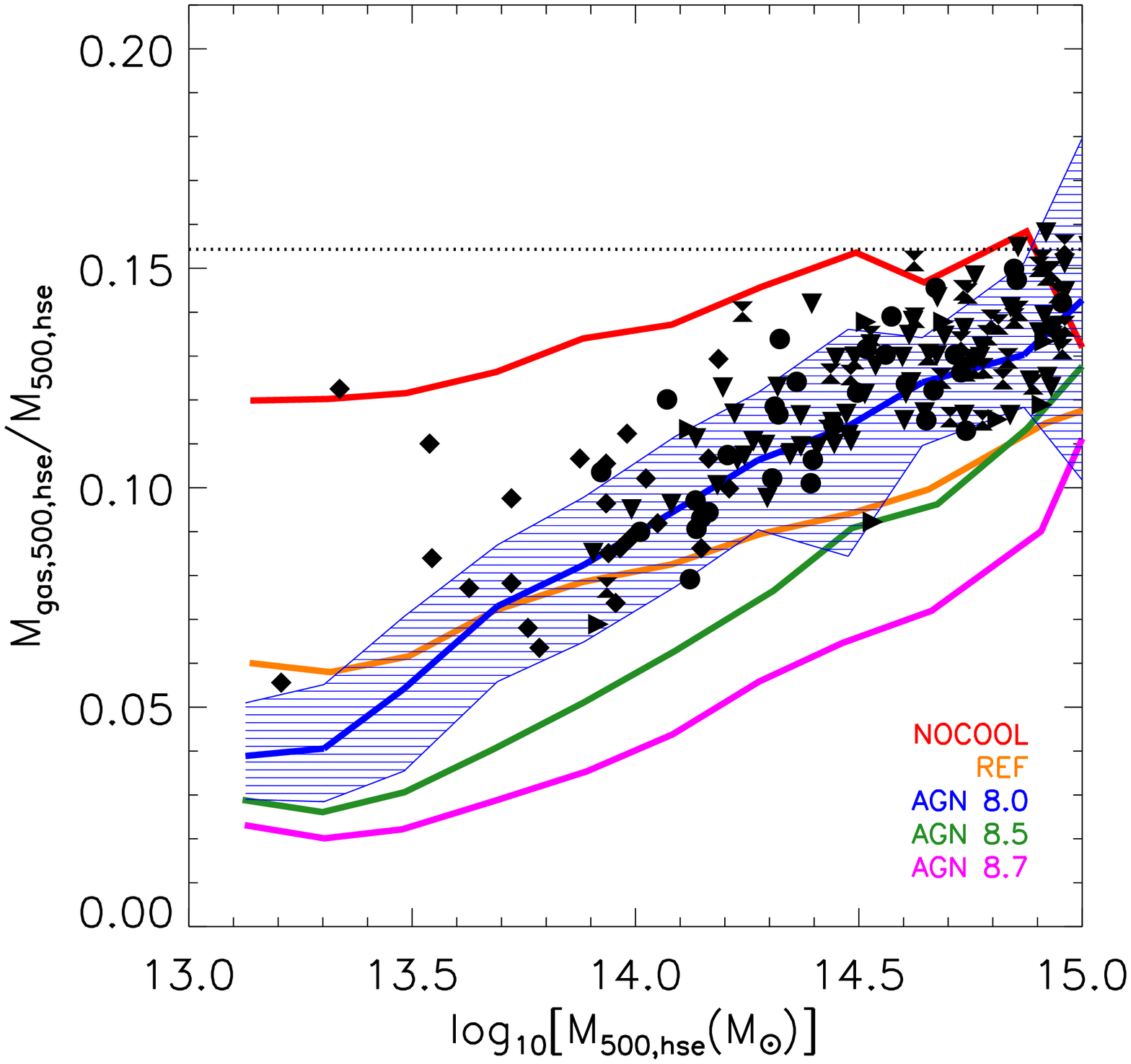}
\caption{The gas mass fraction within $r_{500,hse}$ as a function of $M_{500,hse}$ at $z=0$. The filled black circles (clusters), right-facing triangles (clusters), downward triangles (clusters), hourglass (clusters) and diamonds (groups) represent the observational data of \citet{Pratt2009}, \citet{Vikhlinin2006}, \citet{Lin2012}, \citet{Maughan2008} and \citet{Sun2009}, respectively. The coloured solid curves represent the median gas mass fraction--$M_{500,hse}$ relations in bins of $M_{500,hse}$ for the different simulations and the blue shaded region encloses 68 per cent of the simulated systems for the \agn~8.0 model. The observed trend is reproduced very well by the standard AGN model (\agn~8.0) in the \planck~cosmology (in the \wmap7 cosmology, not shown, it is approximately bracketed by the \agn~8.0 and \agn~8.5 models). Raising the AGN heating temperature further results into too much gas being ejected from (the progenitors of) groups and clusters. The \refsim~model (which lacks AGN feedback) also approximately reproduces the observed trend for low-intermediate masses (though not for $M_{500,hse}\ga10^{14.5}~\textrm{M}_\odot$), but at the expense of significant overcooling (see Fig.~\ref{fig:mass_fstar}).}
\label{fig:mass_fgas}
\end{center}
\end{figure}

\subsubsection{Gas mass fraction--mass relation}

\begin{figure*}
\begin{center}
\includegraphics[width=0.49\hsize]{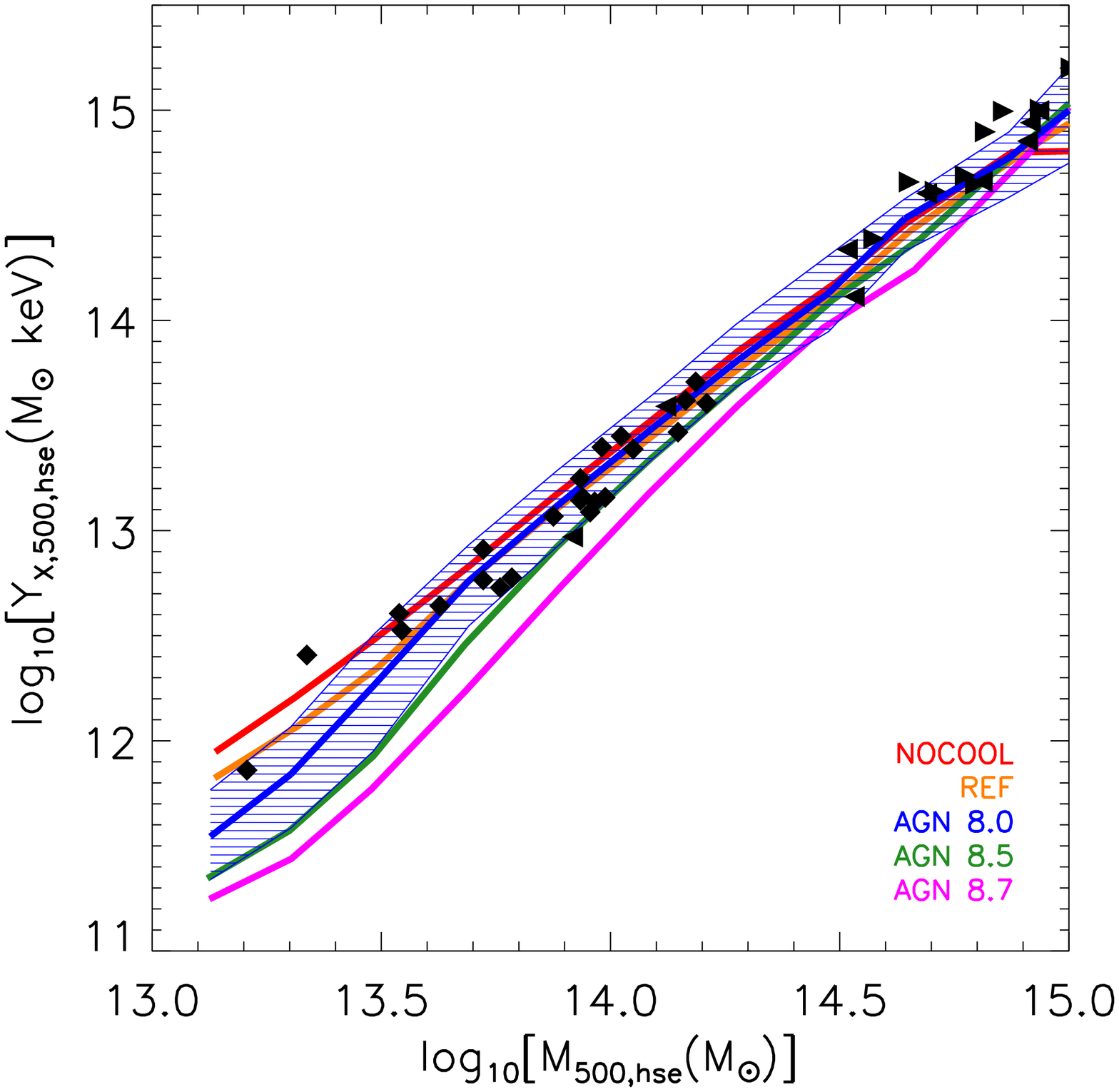}
\includegraphics[width=0.49\hsize]{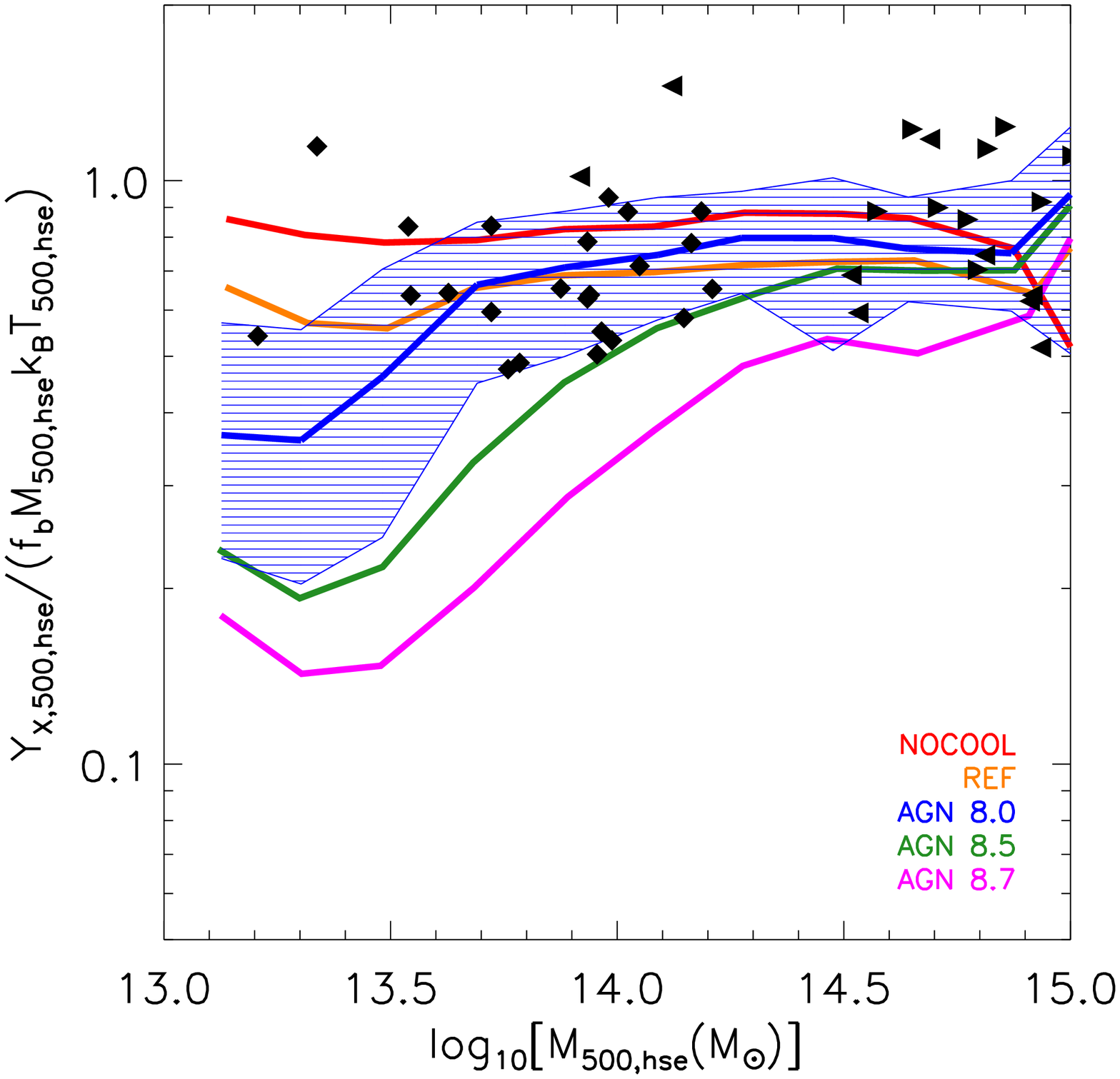}
\caption{The $Y_X-M_{500,hse}$ relation at $z=0$. The filled black left-facing triangles (clusters), right-facing triangles (clusters) and diamonds (groups) represent the observational data of \citet{Vikhlinin2006}, \citeauthor{Planck2012c} and \citet{Sun2009}, respectively. The coloured solid curves represent the median $Y_X$--$M_{500,hse}$ relations in bins of $M_{500,hse}$ for the different simulations and the blue shaded region encloses 68 per cent of the simulated systems for the \agn~8.0 model. In the left panel, we plot the observed $Y_X$ (in $\textrm{M}_\odot$ keV), while in the right panel, $Y_X$ is normalized by $f_b M_{500,hse} k_{B}T_{500,hse}$ to take out the explicit gravitational halo mass dependence. The \agn~8.0 model reproduces the observed trend over approximately two orders of magnitude in mass. Higher heating temperatures result in too low $Y_X$ for low-mass groups relative to the observations (due to over-efficient gas ejection).}
\label{fig:mass_Yx} 
\end{center}
\end{figure*}

In Fig.~\ref{fig:mass_fgas}, we plot the gas mass fraction--$M_{500,hse}$ relation at $z=0$ for the various simulations and compare to observations of individual X-ray-selected systems. The gas mass fraction is measured within $r_{500,hse}$. For the simulated systems, we use our synthetic X-ray observations/analysis methodology to `measure' the halo mass and gas mass fraction of the simulated systems.

As is well known, the observed relation shows a strong trend in gas mass fraction with total system mass, such that galaxy groups have significantly lower fractions compared to massive clusters and the universal baryon fraction $f_b\equiv\Omega_b/\Omega_m$. Some previous observational studies argued that this was due to a much higher star formation efficiency in groups relative to clusters (e.g. \citealt*{Gonzalez2007}; \citealt{Giodini2009}), but some recent observational results suggest that the star formation efficiency of groups is similar to that of clusters \citep[e.g.][]{Leauthaud2012,Budzynski2014} and is therefore far below what is needed to `baryonically close' groups \citep[e.g.][]{Sanderson2013}, even when intracluster light is explicitly accounted for \citep{Budzynski2014}.

The observed trend, as well as its scatter, are reproduced extremely well by the \agn~8.0 model from groups up to massive clusters in the \planck~cosmology. In the \wmap7 cosmology (not shown), which has a universal baryon fraction of $\Omega_b/\Omega_m = 0.167$ (compared to the \planck~value of 0.154 -- dotted horizontal line), the observed trend is approximately bracketed by the \agn~8.0 and \agn~8.5 models (more gas must be ejected in the \wmap7 cosmology to recover the observed gas mass fraction). As demonstrated by \citet{McCarthy2011}, the reduced gas mass fraction with respect to the universal mean in the AGN models is achieved primarily by the ejection of gas from the high redshift progenitors of today's groups and clusters. (Star formation accounts for only $\sim10$ per cent of the removal of hot gas in these models.) The lower binding energies of groups compared to clusters result in more efficient ejection from groups, which naturally leads to the trend in decreasing gas fraction at lower halo masses. This is consistent with the findings of previous simulation studies, such as those of \citet{Bhattacharya2008}, \citet{Puchwein2008}, \citet{Short2009}, \citet{Fabjan2010}, \citet{Stanek2010} and \citet{Planelles2013}. 

Note that increasing the heating temperature of the AGN further results in too much gas being ejected from all systems. The \refsim~model, which lacks AGN feedback altogether, also yields reasonable gas mass fractions, but the relation with mass is flatter than observed, because the star formation efficiency does not depend strongly on halo mass. The low gas fractions in this model are achieved by overly efficient star formation (see Fig.~\ref{fig:mass_fstar}). 

We note that the non-radiative run, \nocool, has a slight trend with mass and that some massive clusters apparently have gas mass fractions well in excess of the universal baryon fraction (the scatter, not shown, is somewhat larger in magnitude compared to that of the \agn~8.0 model). Naively, this would appear to contradict previous studies which also examined non-radiative simulations and found that the baryon fraction does not depend on halo mass and is very nearly the universal fraction within $r_{500}$ with little scatter \citep[e.g.][]{Crain2007}. There is, in fact, no contradiction -- our non-radiative results agree very well with previously studies when considering the {\it true} baryon fraction vs halo mass trend. The slight trend indicated in Fig.~\ref{fig:mass_fgas} and the large scatter (not shown) are due to biases in the recovered gas density and total mass profiles introduced during the synthetic X-ray observation analysis. In particular, because it is unable to cool, there is a lot more gas at short cooling times (low temperature and high density) in this run, which biases the recovered ICM density and temperature due to its high X-ray emissivity. These biases are significantly reduced in radiative simulations, where cooling and feedback tend to remove low-entropy gas from the systems. 

\subsubsection{$Y_X-$mass relation} 
\label{sec:mass_Yx}

\begin{figure*}
\begin{center}
\includegraphics[width=0.49\hsize]{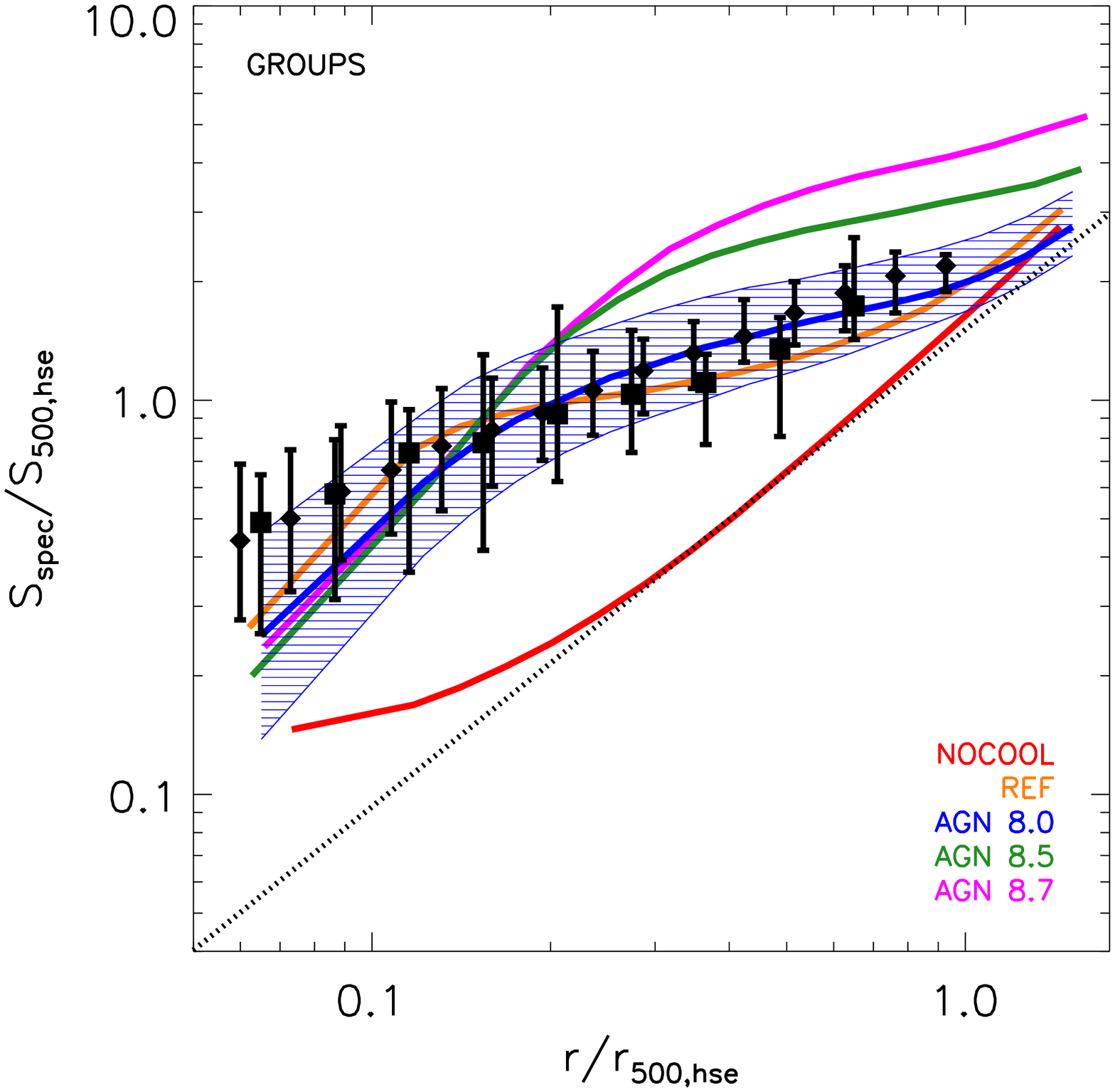}
\includegraphics[width=0.49\hsize]{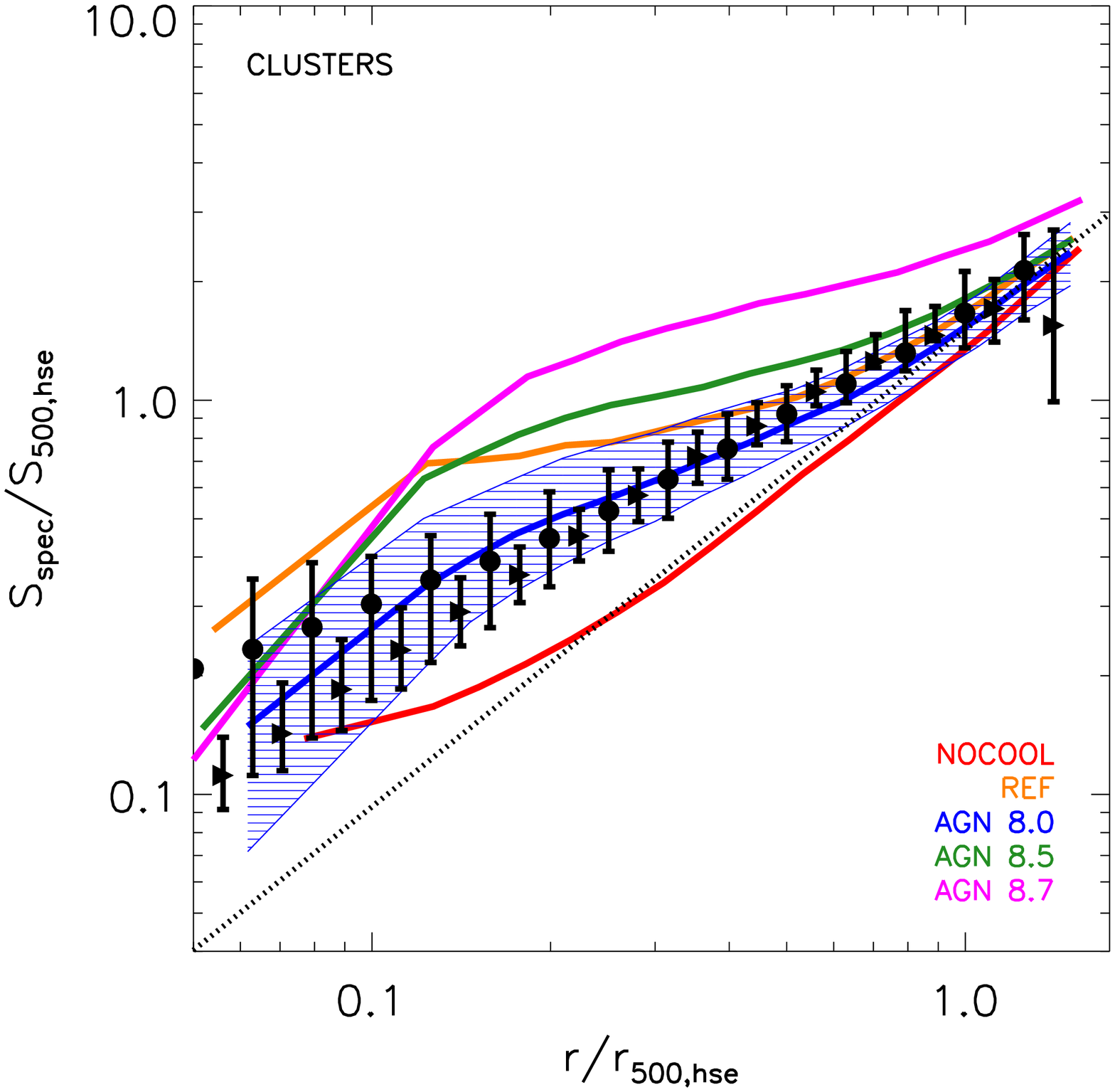}
\caption{The radial entropy profiles of groups (\emph{left}) and clusters (\emph{right}) at $z=0$. The simulated systems have been selected to match the median mass of the observational data. The filled black diamonds (groups), squares (groups), circles (clusters) and right-facing triangles (clusters) with error bars correspond to the observational data of \citet{Sun2009}, \citet*{Johnson2009}, \citet{Pratt2010} and \citet{Vikhlinin2006} (in the latter case, the entropy profiles were obtained by combining their best-fitting density and temperature profiles), respectively. The error bars enclose 90 per cent and 68 per cent of the observed systems for groups and clusters, respectively. The dotted line represents the power law fit of \citet{Voit2005a} to the entropy profiles of a sample of simulated non-radiative SPH groups and clusters. The coloured solid curves represent the median entropy profiles for the different simulations and the blue shaded region encloses 68 per cent of the simulated systems for the \agn~8.0 model. The standard \agn~8.0 model reproduces the observed radial profiles of groups and clusters over 1.5 decades in radius, and the observed scatter is also broadly reproduced.}
\label{fig:Sprof}
\end{center}
\end{figure*}

In the left panel of Fig.~\ref{fig:mass_Yx}, we plot the $Y_X-M_{500,hse}$ relation at $z=0$ for the various simulations and compare to observations of individual X-ray-selected systems. $Y_X$ is the X-ray analogue of the SZ flux and is hence defined as the product of the hot gas mass within $r_{500,hse}$ and the core-excised mean X-ray spectral temperature (as in Fig.~\ref{fig:mass_Tx}) and is thus closely related to the total thermal energy of the ICM. \citet*{Kravtsov2006} first proposed $Y_X$ as a cluster mass proxy, arguing that it should be relatively insensitive to the details of ICM physics and merging.

In the left panel of Fig.~\ref{fig:mass_Yx}, we see that the various simulations indeed yield similar $Y_X-M_{500,hse}$ relations (the \refsim, \nocool, and \agn~8.0 models reproduce the data best) and $Y_X$ is clearly strongly correlated with system mass. However, due to the large dynamic range in $Y_X$ plotted in the left panel of Fig.~\ref{fig:mass_Yx}, one perhaps gets a misleading impression of the sensitivity of $Y_X$ to ICM physics. To address this, we plot in the right panel of Fig.~\ref{fig:mass_Yx} the dimensionless quantity $Y_X / (f_b M_{500,hse} k_{B}T_{500,hse})$, where $k_{B}T_{500,hse}\equiv\mu m_{p} G M_{500,hse} / 2 r_{500,hse}$. The denominator takes out the explicit halo mass dependence of $Y_X$ and greatly reduces the dynamic range on the y-axis, allowing for a better examination of the sensitivity of $Y_X$ to the important non-gravitational physics. Note that $f_b M_{500,hse} k_{B}T_{500,hse}$ is the $Y_X$ a cluster of mass $M_{500,hse}$ would possess if the hot gas were isothermal with the virial temperature and the gas mass fraction had the universal value (i.e. the self-similar prediction).

From the right hand panel of Fig.~\ref{fig:mass_Yx}, one immediately concludes that $Y_X$ \emph{is} in fact sensitive to ICM physics, contrary to the claims of \citet{Kravtsov2006}. More specifically, energetic AGN, which were not examined by Kravtsov et al., can eject large quantities of gas that can significantly lower $Y_X$. This reduction in gas mass can be compensated to a degree by the slight increase in temperature due to the fact that much of the ejected gas had low entropy (and also additional high entropy gas is able to accrete within $r_{500}$; \citealt{McCarthy2011}). However, hydrostatic equilibrium forces the temperature of the ICM to remain near the virial temperature, and thus arbitrarily large amounts of gas ejection cannot be compensated for.

At $z=0$, observed groups and clusters have sufficiently high gas mass fractions that $Y_X$ is not significantly depressed compared to the self-similar prediction. However, Fig.~\ref{fig:mass_Yx} should serve as a warning against blindly applying $Y_X$ to, e.g. lower halo masses and/or higher redshifts, where independent direct halo mass estimates are increasingly scarce. This caution should also obviously be heeded (perhaps even more so) by studies which use gas mass (fractions) as total mass proxies as opposed to $Y_X$.

%% Profiles

\subsection{Profiles}
\label{sec:prof}

\subsubsection{Entropy}

\begin{figure}
\begin{center}
\includegraphics[height=0.75\hsize]{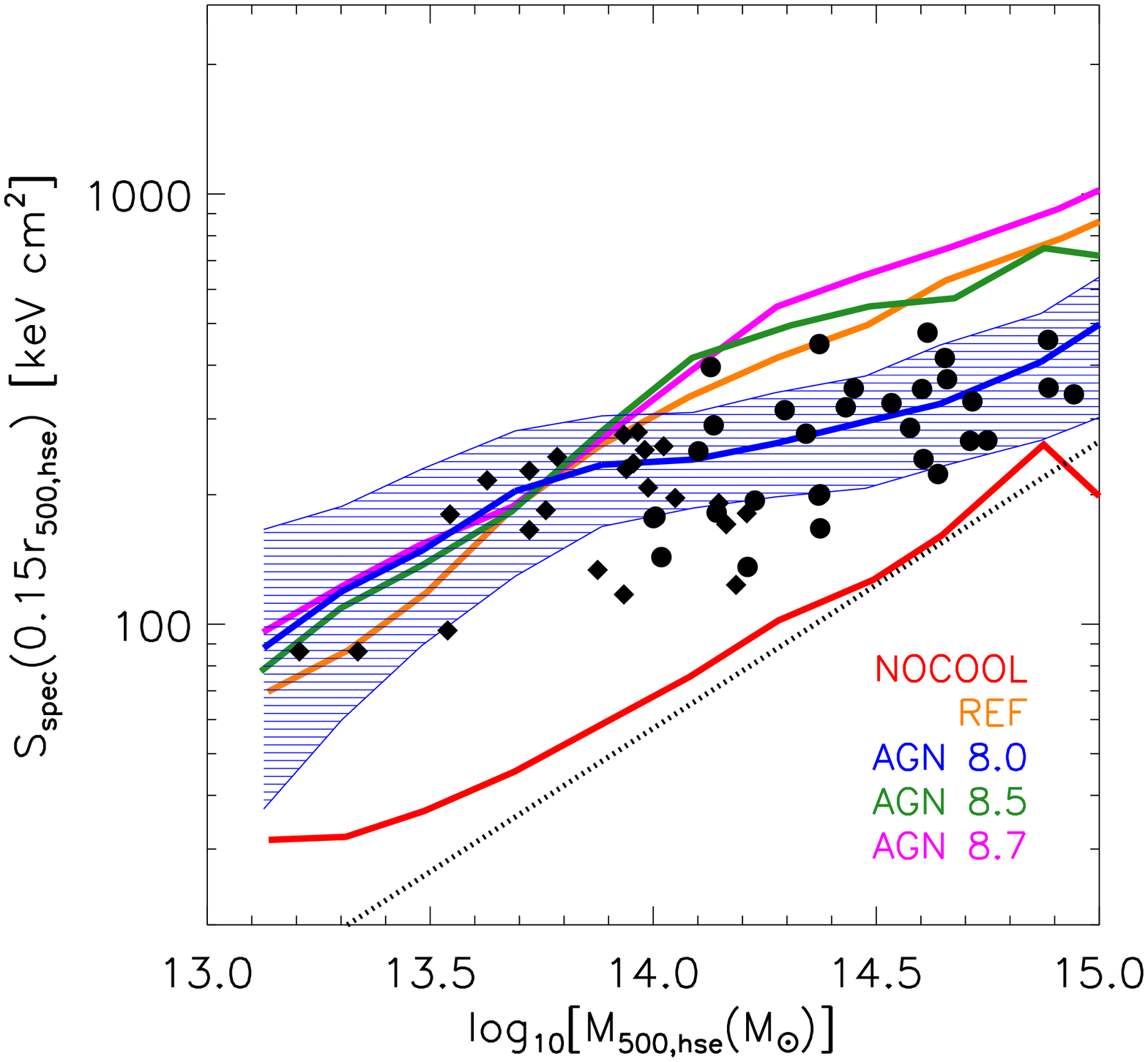}
\includegraphics[height=0.75\hsize]{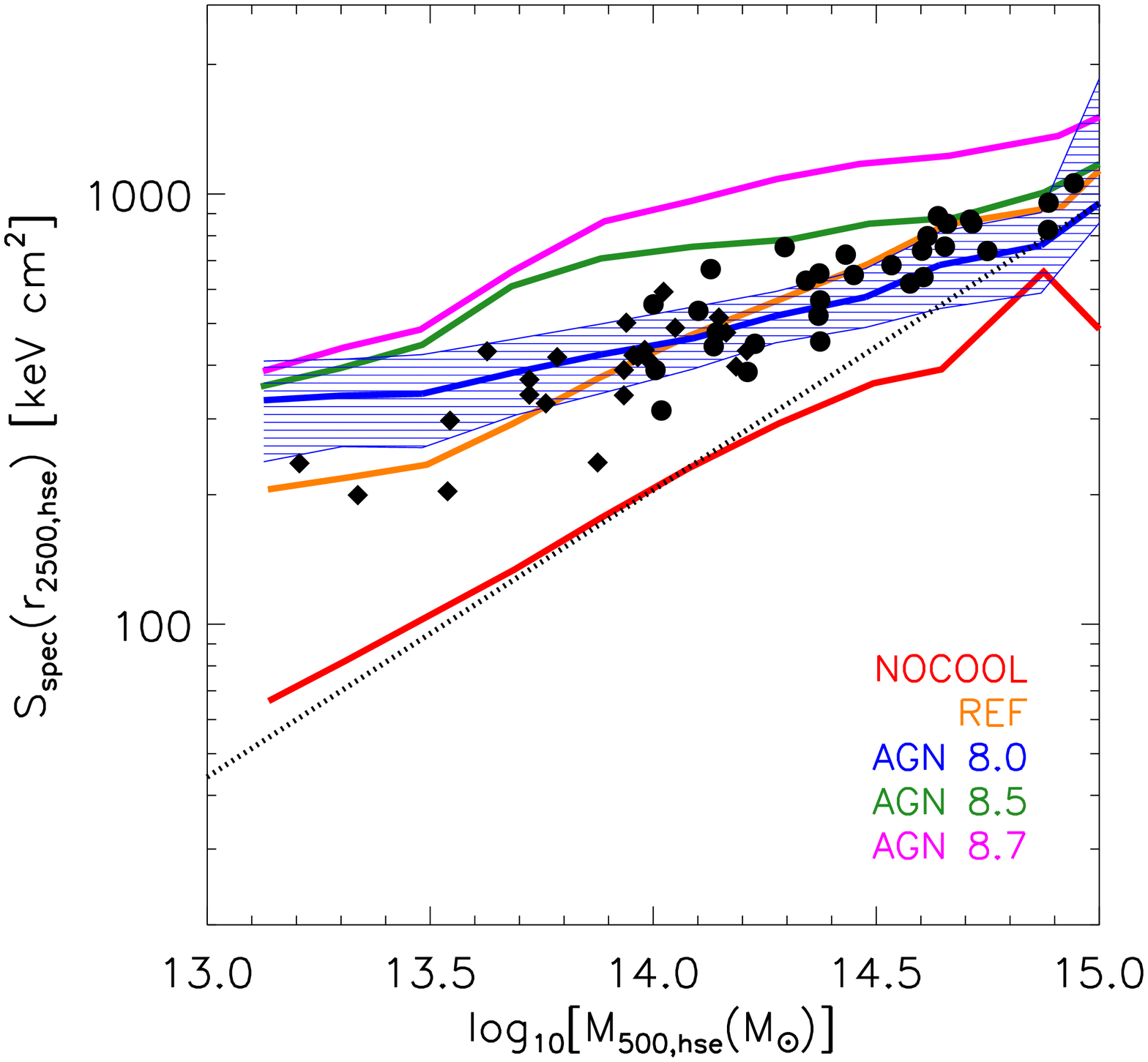}
\includegraphics[height=0.75\hsize]{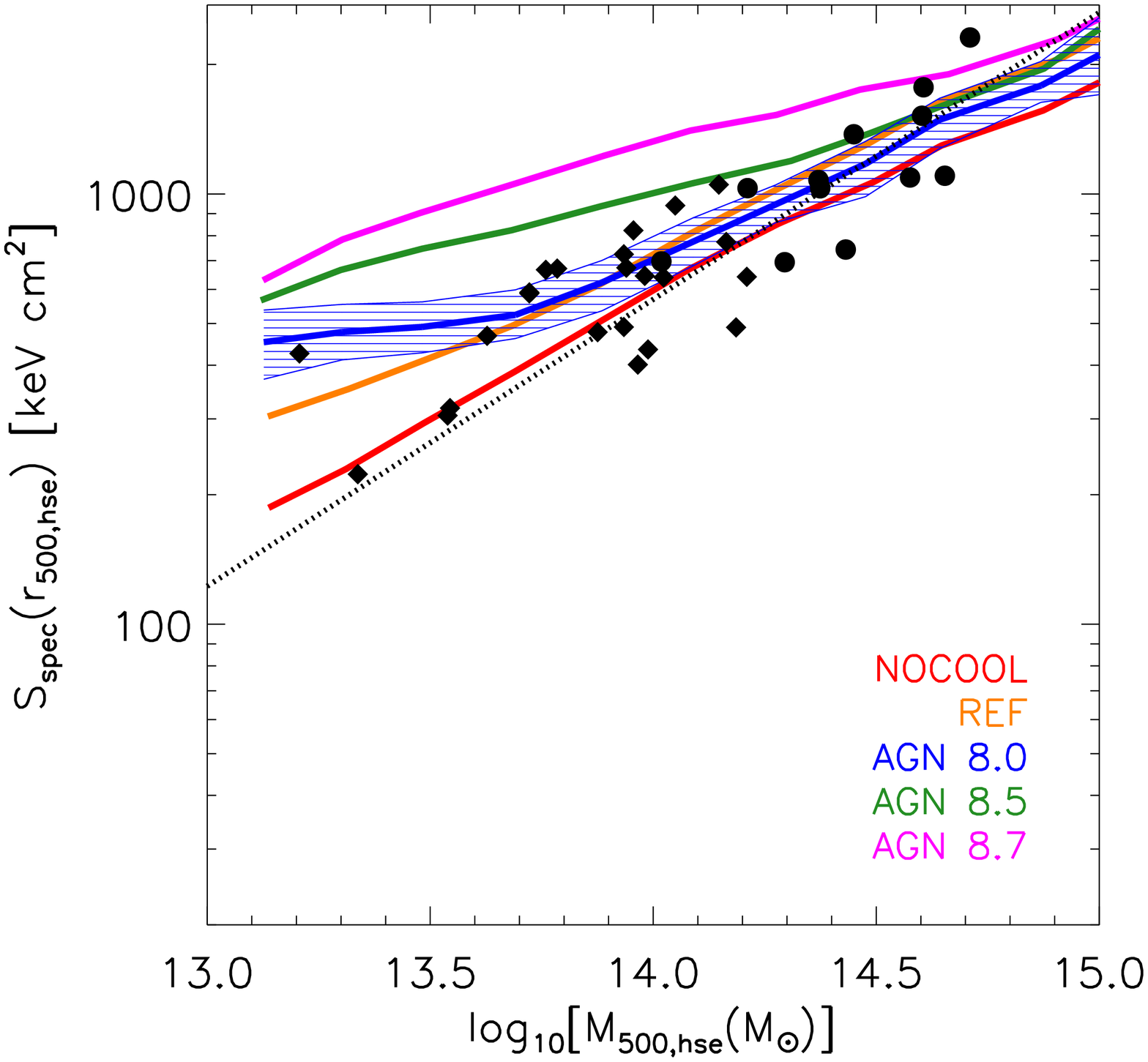}
\caption{The $z=0$ entropy measured at various characteristic radii ($0.15r_{500,hse}$ (\emph{top}), $r_{2500,hse}$ (\emph{middle}) and at $r_{500,hse}$ (\emph{bottom})) as a function of $M_{500,hse}$. The filled black circles (clusters) and diamonds (groups) correspond to the observations of \citet{Pratt2010} and \citet{Sun2009}, respectively, while the dotted line represents the power law fit of \citet{Voit2005a} to the entropy profiles of a sample of simulated non-radiative SPH groups and clusters. The coloured solid curves represent the median reference entropy--mass relations in bins of $M_{500,hse}$ for the different simulations and the blue shaded region encloses 68 per cent of the simulated systems for the \agn~8.0 model. Deviations from the self-similar prediction are largest at small radii and low halo masses. Only the \agn~8.0 model reproduces the observational data over the full range of radii and halo masses.}
\label{fig:Sref}
\end{center}
\end{figure}

In Fig.~\ref{fig:Sprof}, we plot the three-dimensional radial entropy profiles of groups (left panel) and clusters (right panel) for the various simulations and compare to observations of X-ray-selected systems. As the shape and amplitude of the entropy profiles are fairly strong functions of halo mass (as shown in Fig. ~\ref{fig:Sref}), we have slightly re-sampled the mass distributions of the observational and simulated samples so that they have approximately the same median mass for both, which is $M_{500} \approx 8.6\times10^{13}~\textrm{M}_{\odot}$ for groups and $M_{500} \approx 3.5\times10^{14}~\textrm{M}_{\odot}$ for clusters. (This was achieved by keeping only the simulated groups with $5.75\times10^{13}~\textrm{M}_\odot \le M_{500} \le 1.54\times10^{14}~\textrm{M}_\odot$, the simulated clusters with $2.5\times10^{14}~\textrm{M}_\odot \le M_{500} \le 10^{15}~\textrm{M}_\odot$, the \rexcess~clusters with $M_{500}\ge1.5\times10^{14}~\textrm{M}_\odot$ and the \citet{Vikhlinin2006} clusters with $1.2\times10^{14}~\textrm{M}_\odot \le M_{500} \le 1.1\times10^{15}~\textrm{M}_\odot$.)
We use the definition of entropy commonly used in X-ray astronomy, i.e. $S\equiv k_B T/n_{e}^{2/3}$, which here has units of $\textrm{keV cm}^2$ and is related to the thermodynamic entropy by a logarithm and an additive constant.
We normalize the radii by $r_{500,hse}$ and the entropies by the characteristic entropy scale $S_{500,hse}$, which is defined as
\begin{equation}
S_{500,hse}\equiv \frac{k_B T_{500,hse}}{n_{e,500,hse}^{2/3}}=\frac{GM_{500,hse}\mu m_p}{2r_{500,hse}(500f_b \rho_{crit}/(\mu_{e}m_p))^{2/3}},
\end{equation}
where $\mu_e$ is the mean molecular weight per free electron, in order to take out the explicit halo mass dependence. We also show the baseline entropy profile of \citet*{Voit2005a} as a dotted line on both panels. This represents the self-similar answer, which was obtained by fitting a power-law to the entropy profiles of a sample of non-radiative SPH groups and clusters. Finally, as the observed entropy profiles were obtained through spectral fitting of X-ray observations, we have used our synthetic X-ray observations methodology to compute spectral entropy profiles for the simulated systems. 

As is well known, observed groups and clusters exhibit a significant level of `excess entropy' compared to the self-similar expectation \citep*[e.g.][]{Ponman1999,Ponman2003}, which is a clear signature of the non-gravitational physics of structure formation. This effect is stronger in groups compared to clusters. Fig.~\ref{fig:Sprof} shows that all the radiative models (\refsim~and the AGN models) yield profiles that are similar to the observed ones in the central regions ($r \la 0.2r_{500,hse}$) of groups. In more massive clusters, however, only the \agn~8.0 model provides an adequate match to the observations. At intermediate/large radii, the \agn~models with the two highest heating temperatures (\agn~8.5 and \agn~8.7) have too high entropy at intermediate and large radii compared to the observed levels (particularly in groups), due to the ejection of too much (preferentially low-entropy) gas from the progenitors of the present-day systems. \citet{Short2013} also find that the inclusion of AGN feedback leads to better agreement at intermediate radii for clusters.

\begin{figure*}
\begin{center}
\includegraphics[width=0.49\hsize]{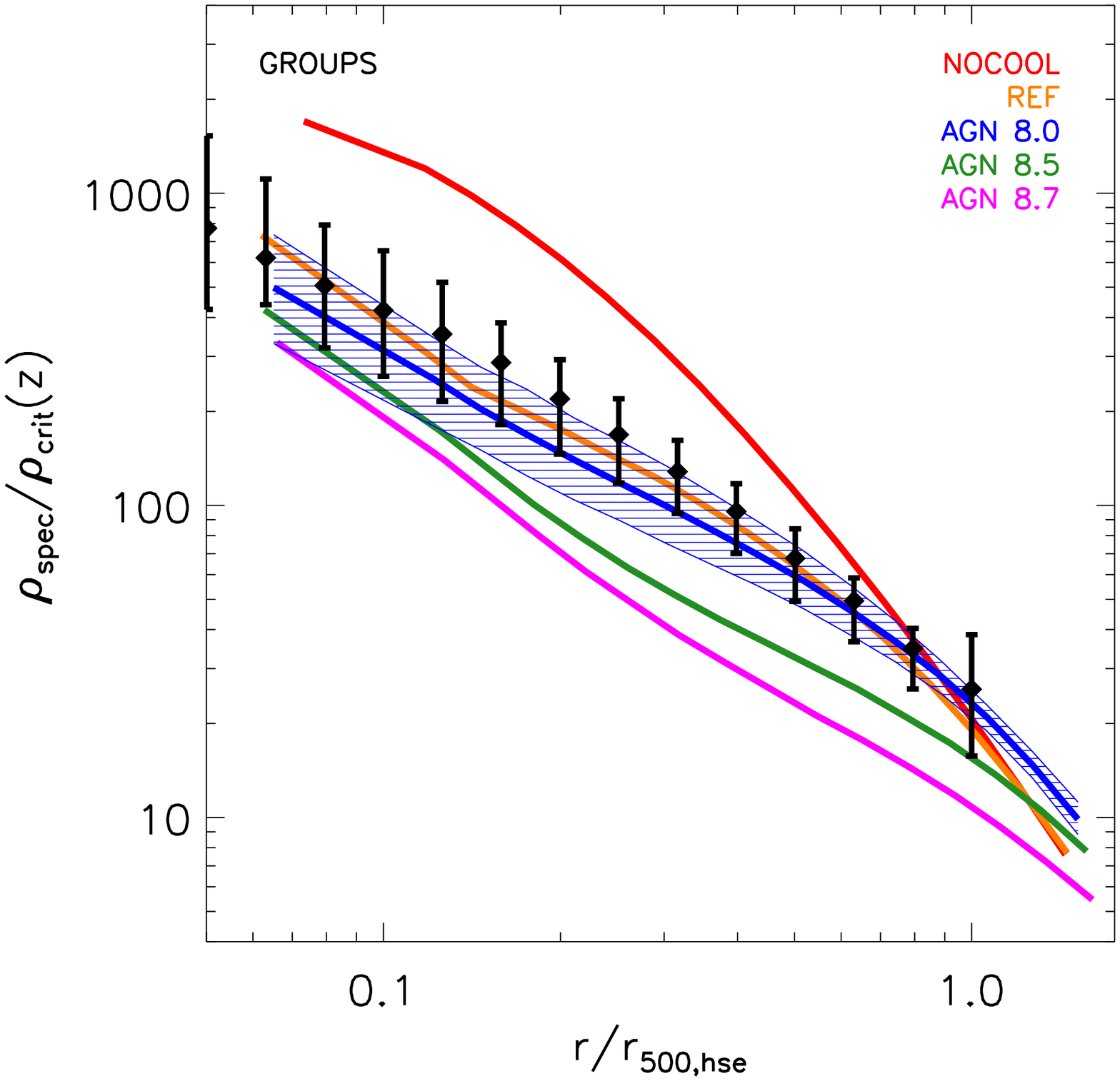}
\includegraphics[width=0.49\hsize]{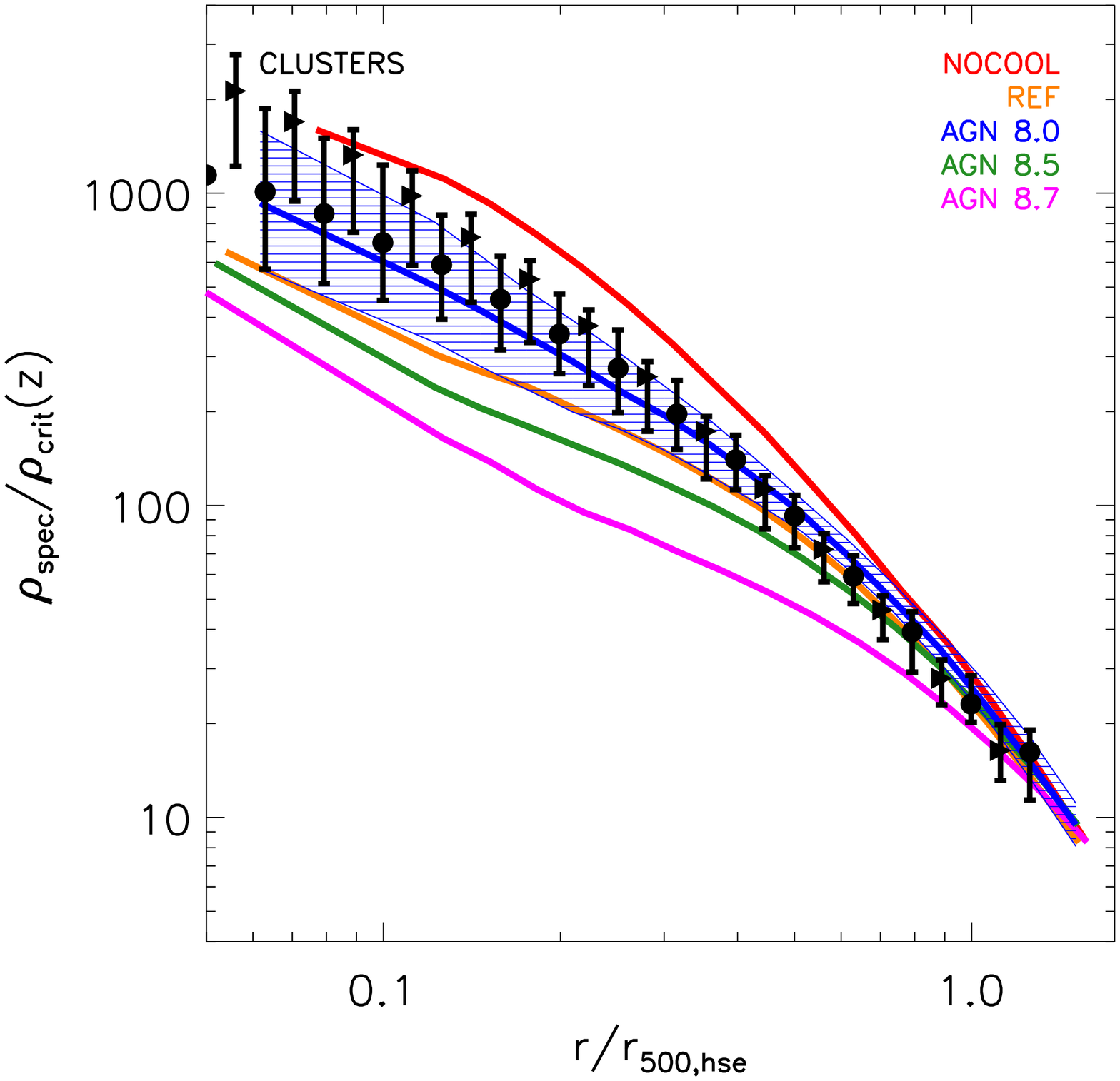}
\caption{The radial density profiles of groups (\emph{left}) and clusters (\emph{right}) at $z=0$. The simulated systems have been selected to match the median mass of the observational data. The filled black diamonds (groups), circles (clusters) and right-facing triangles (clusters) with error bars correspond to the observational data of \citet{Sun2009}, \citet{Croston2008} and \citet{Vikhlinin2006}, respectively. The error bars enclose 68 per cent of the observed systems. The coloured solid curves represent the median density profiles for the different simulations and the blue shaded region encloses 68 per cent of the simulated systems for the \agn~8.0 model. The observed trends are reproduced well in the \planck~cosmology by the standard AGN model (\agn~8.0). In the \wmap7 cosmology (not shown), the simulated density profiles are shifted up by approximately 10 per cent.}
\label{fig:rhoprof}
\end{center}
\end{figure*}

We note that the consequences of observational selection are also apparent in Fig.~\ref{fig:Sprof}. In particular, the filled black circles in the right panel represent the median entropy profile from \citet{Pratt2010}, derived from \rexcess~-- a representative sample of 33 clusters derived from a flux-limited parent sample \citep{Bohringer2007}, whereas the black right-facing triangles represent the sample of \citet{Vikhlinin2006}, who targeted relaxed, cool core clusters. It is apparent that the clusters from the \citet{Pratt2010} sample have a higher mean central entropy and larger central scatter, as one might expect, since there is no requirement for their clusters to have a central temperature dip (which necessitates a low central entropy). The comparison to the \citet{Pratt2010} sample is therefore perhaps more appropriate. However, there still remains the question of how `representative' flux-limited samples really are relative to a halo mass-selected sample, as typically derived from models/simulations such as those presented here. While it is doubtful that X-ray surveys are missing many massive nearby clusters, it is nevertheless possible that the mix of clusters in a given bin may be skewed. Furthermore, our confidence in the completeness of X-ray surveys (even above a given luminosity, let alone mass) weakens considerably as we move into the group regime.

To better explore the relatively strong dependence on halo mass apparent in Fig.~\ref{fig:Sprof}, we plot in Fig.~\ref{fig:Sref} the entropy at three reference radii ($0.15r_{500,hse}\approx r_{13000,hse}$, $r_{2500,hse}\approx0.45r_{500,hse}$ and $r_{500,hse}$ from top to bottom) as a function of $M_{500,hse}$ for the various simulations and compare to observations of individual X-ray-selected groups and clusters. We also show the baseline entropy profile of \citet{Voit2005a} as a dotted line in all three panels. Deviations from the baseline self-similar results are strongest at the lowest halo masses and smallest radii. Only the standard AGN model (\agn~8.0) is able to reproduce the observed trends with radius and halo mass. Similar results were obtained by \citet{Fabjan2010} and \citet{Planelles2014}, but they only looked at the relation for the largest two of the characteristic radii. 

\subsubsection{Density}
\label{sec:rhoprof}

In Fig.~\ref{fig:rhoprof}, we plot the three-dimensional radial density profiles of groups (left panel) and clusters (right panel) for the various simulations and compare to observations of X-ray-selected systems (symbols with error bars). As we did for the entropy profile comparison above, we have approximately matched the median masses of the observed and simulated samples by excising some systems from each. The resulting samples are identical to those used for the entropy profiles in the previous subsection. We normalize the radii by $r_{500,hse}$ and the densities by the critical density of the universe for our adopted cosmological parameters. Finally, as the observed density profiles were obtained through spectral fitting of X-ray observations, we have used our synthetic X-ray observations methodology to compute spectral density profiles for the simulated systems. 

The \agn~8.0 model reproduces the observed profiles (including the scatter) quite well over the whole radial range for both groups and clusters in the \planck~cosmology. (In the \wmap7 cosmology, the simulation gas density profiles are shifted up by approximately the ratio of universal baryons in \wmap7 and \planck~cosmologies.) Increased heating temperatures, which lead to more violent and bursty AGN feedback (e.g. \agn~8.7), result in a strongly reduced density, especially in the central regions and in low-mass systems. Conversely, when both feedback and radiative cooling are omitted (\nocool), the gas is too dense and too centrally concentrated. It is worth noting that the non-gravitational physics of galaxy formation has a noticeable effect on the group gas density profiles as far out as $\sim r_{500,hse}$, whereas in the case of clusters, the profiles have all approximately converged to the self-similar answer at these radii.

As discussed above, the role of observational selection is an important caveat to bear in mind, particularly for groups. Note that the median central density of the observed sample of groups in \citet{Sun2009} is slightly higher than that of our fiducial AGN model, consistent with the offset in the mass-luminosity relation at low masses (see Fig.~\ref{fig:mass_Lx}). As we discussed in Section~\ref{sec:mass_Lx}, however, the \citet{Sun2009} sample has a higher mean X-ray luminosity compared to other observational group samples, most likely due to selection.

\begin{figure}
\begin{center}
\includegraphics[width=1.0\hsize]{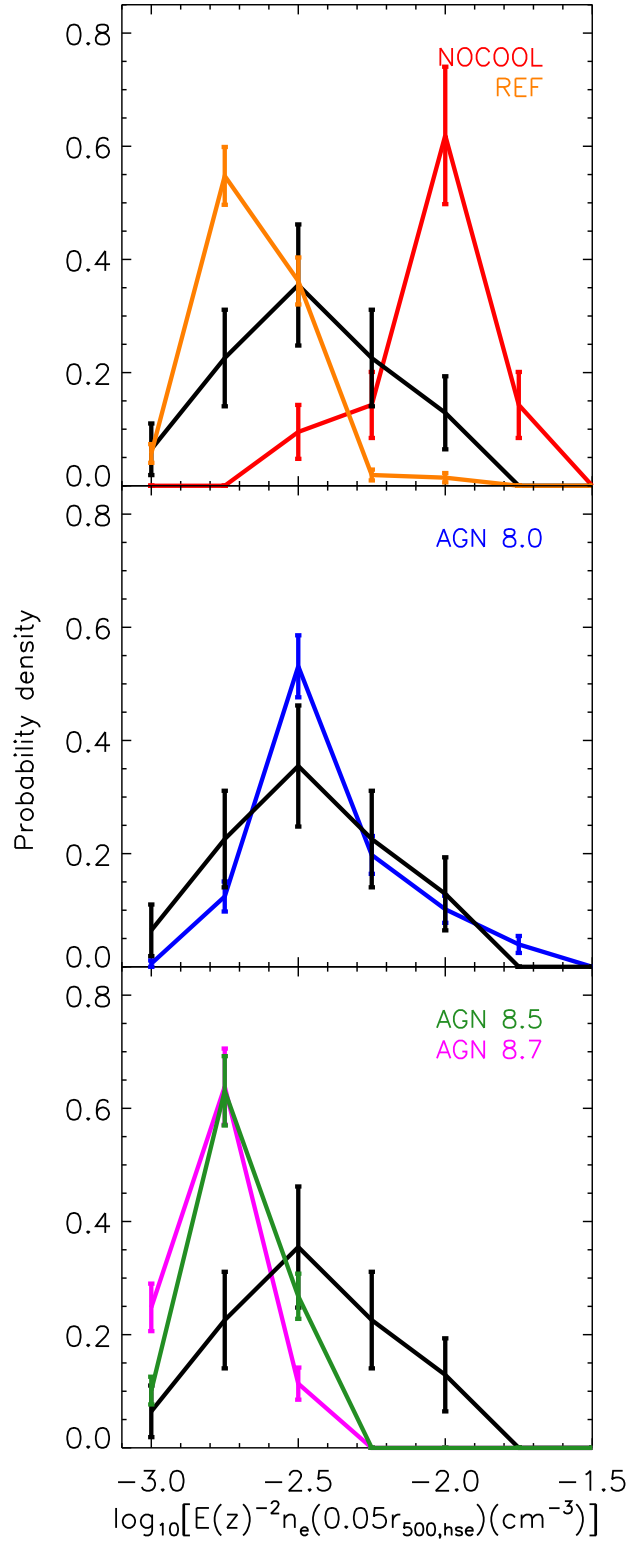}
\caption{Distribution of central (at $0.05r_{500,hse}$) electron densities at $z=0$. The thick solid histograms (red, orange, blue, green and magenta) are for the different simulations while the black one corresponds to the observational data of \citet{Croston2008} with $z\le0.25$ scaled to $z=0$ assuming self-similar evolution. The error bars represent Poisson noise.  The \agn~8.0 model reproduces the observed large spread in the central density distribution of the hot gas, which shows no strong evidence for bimodality.}
\label{fig:n0}
\end{center}
\end{figure}

\subsubsection{Demographics of cluster cores}
\label{sec:n0}

The observed large scatter in the properties of the hot gas in the cores of galaxy clusters is a subject that has attracted much interest in recent years. It was previously noted that the scatter in the observed global scaling relations, such as the luminosity--temperature relation, is driven primarily by the scatter in the thermodynamic properties of the gas within the central $\sim 200$ kpc (e.g. \citealt{Fabian1994,McCarthy2004,McCarthy2008}). The origin of this scatter is still being debated. It may be due to merger activity and/or differences in the feedback histories of clusters. It is of interest to see whether the simulations presented here reproduce the detailed scatter at small radii. 

Detailed studies of the radial structure of the gas with {\it Chandra} and {\it XMM-Newton} have suggested that there may be a bimodality in the central entropy \citep{Cavagnolo2009,Pratt2010}, although this has been called into question recently \citep*{Panagoulia2014}. As pointed out by \citet{Panagoulia2014}, the derived central entropy is sensitive to what is assumed about the temperature distribution at small radii, which cannot be measured in as finely spaced radial bins as the gas density and is somewhat sensitive to the uncertain metallicity of the gas. Furthermore, by experimentation, we have found the results to be sensitive to the way in which the data is binned in radius when fitting power-law + constant models to the entropy distribution (as done in the Cavagnolo et al.\ and Pratt et al.\ studies). 

To overcome these issues, we adopt a non-parametric approach applied to the central gas density distribution, which can be robustly determined from observations. In particular, we plot the gas density measured at $0.05 r_{500,hse}$ in Fig.~\ref{fig:n0} and observational estimates of \citet{Croston2008} for the representative \rexcess~cluster sample. As in previous plots, we re-sample the mass distributions to achieve the same median mass for the observed and simulated samples.

Encouragingly, the fiducial AGN model has a central density distribution that is quite similar to the observed one.  The central density varies by over an order of magnitude in both. Furthermore, we see no strong evidence for a bimodal distribution in either the observed or simulated density distributions.  This does not necessarily imply that the entropy will not be bimodal, as the entropy depends on the temperature as well. Note that to have a bimodal distribution in the entropy but not in the density requires there to be a bimodal distribution in the shape of the potential well at small radii (or else the system is not convectively stable), with high-entropy systems having {\it deeper} potential wells. In our models, however, the entropy measured at $0.05r_{500,hse}$ is {\it not} bimodal, in qualitative agreement with the recent observational findings of \citet{Panagoulia2014}.

Based on the above, the dividing line between `cool core' and `non-cool core' is therefore somewhat arbitrary. The fact that the fiducial AGN model has a similar central density distribution to that of the \rexcess~sample implies that, regardless of how they are exactly defined, both types of clusters are present in this model and in approximately the correct proportion. 

Given that the fiducial AGN model reproduces the observed core demographics rather well, we intend to address the origin of the scatter in the simulations in a future study.

%%%%%%%%%%%%%%%%%
% Sunyaev-Zel'dovich scalings %
%%%%%%%%%%%%%%%%%

\section{Sunyaev--Zel'dovich scalings}

\begin{figure*}
\begin{center}
\includegraphics[width=0.49\hsize]{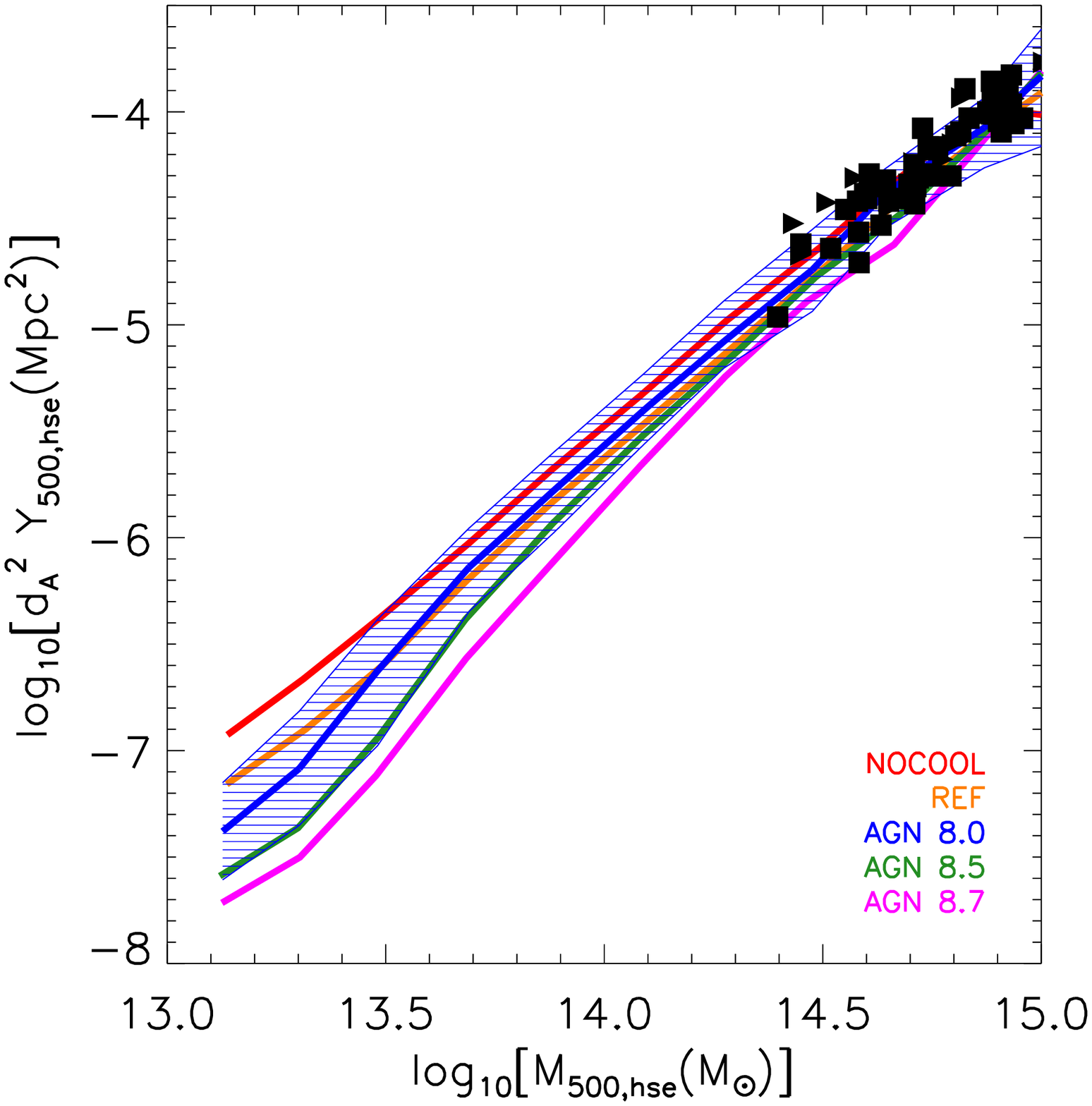}
\includegraphics[width=0.49\hsize]{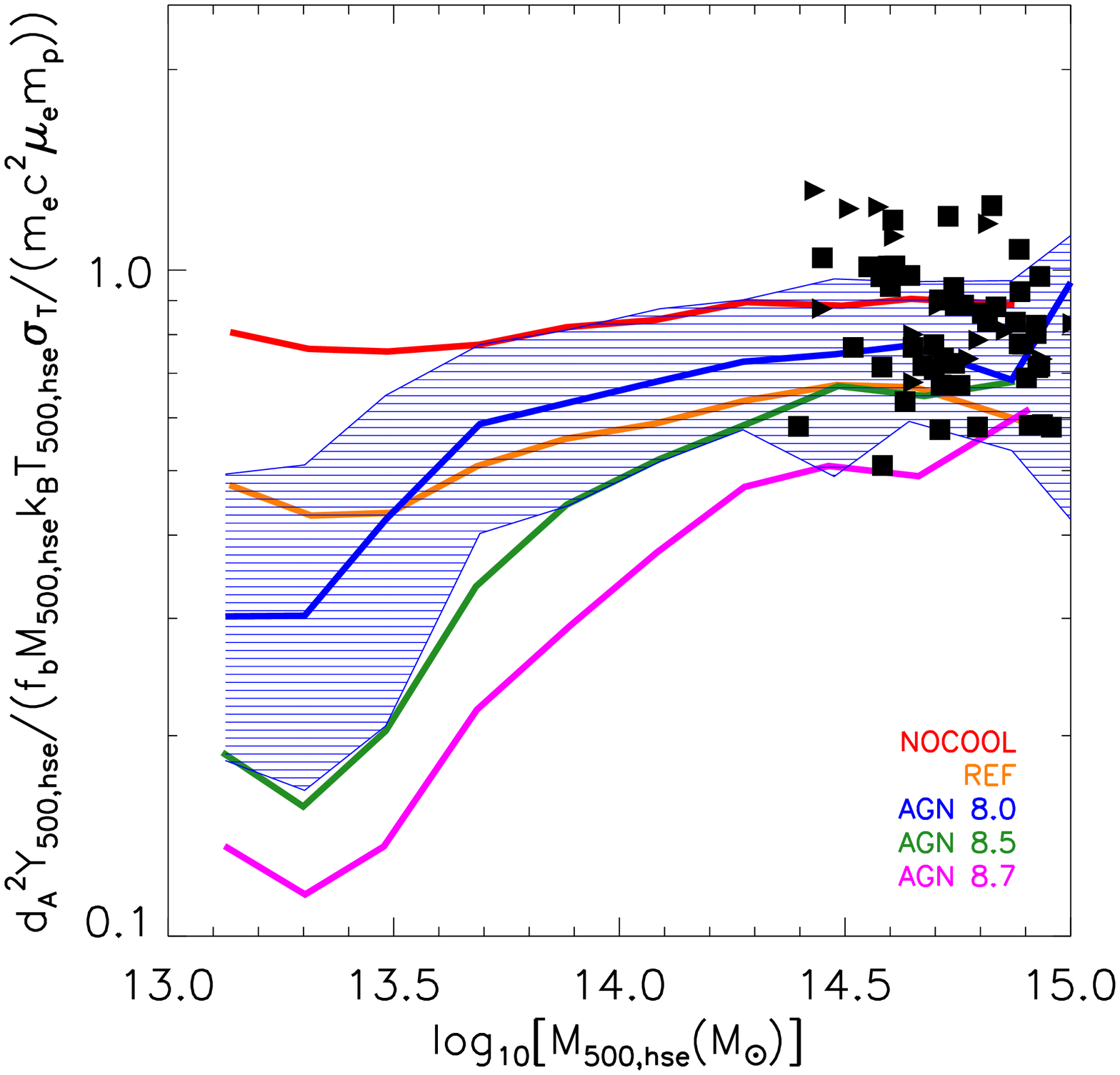}
\caption{The $Y_{500,hse}-M_{500,hse}$ relation at $z=0$. The filled black squares and right-facing triangles represent the observational data of \citetalias{Planck2011a} and \citetalias{Planck2012a} with $z\le 0.25$, respectively. The solid curves (red, orange, blue, green and magenta) represent the median SZ flux--$M_{500,hse}$ relations in bins of $M_{500,hse}$ for the different simulations and the blue shaded region encloses 68 per cent of the simulated systems for the \agn~8.0 model. In the left panel, we plot the observed SZ signal (in $\textrm{Mpc}^2$), whereas in the right panel, the SZ flux is normalized by $\sigma_T/(m_e c^2 \mu_e m_p) f_b M_{500,hse} k_B T_{500,hse}$ in order to take out the explicit gravitational mass dependence. Consistent with the conclusions derived from the X-ray comparisons in Section~\ref{sec:Xrayprops}, the fiducial AGN model (\agn~8.0) reproduces the observed trend well in the \planck~cosmology. In the \wmap7 cosmology (not shown), the simulated curves are shifted up by approximately 10 per cent, so that more gas ejection (a slightly higher heating temperature) is required to reproduce the normalization.} 
\label{fig:mass_Y500}
\end{center}
\end{figure*}

The Sunyaev--Zel'dovich effect provides an alternative, complementary way to probe the thermodynamic state of the hot gas in groups and clusters (see e.g. \citealt*{Birkinshaw1999,Carlstrom2002} for reviews). Below we compare the simulated and observed integrated SZ fluxes as a function of halo mass. Integrated over the volume of a system, the SZ effect is proportional to the total thermal energy content of the hot gas.

In Fig.~\ref{fig:mass_Y500}, we plot the SZ flux--$M_{500,hse}$ relation for the various simulations and compare to observations of individual SZ selected systems (re-)discovered by the \planck~satellite, mostly during the first ten months of its mission (the Early Sunyaev--Zel'dovich catalogue; \citeauthor{Planck2011a}) and, either followed up in X-ray with \xmm~(\citeauthor{Planck2011b,Planck2012a,Planck2012d}) using Director's Discretionary Time, or with high-quality archival \xmm~data (\citeauthor{Planck2011c,Planck2012c}). As we are comparing the observational data to the $z=0$ simulation results, we have only kept the \planck~systems with $z\le0.25$. Since the observational mass measurements (and apertures within which the SZ fluxes are measured) are based on either a hydrostatic analysis of the X-ray observations, or on the \citet{Arnaud2010} $Y_{X}-M_{500,hse}$ relation which was calibrated using a sample of 20 nearby relaxed clusters with high quality \xmm~X-ray data\footnote{Eight clusters come from the sample of \citet*{Arnaud2007} and the remaining 12 are relaxed \rexcess~clusters with mass profiles measured at least out to $R_{550}$.}, we use the hydrostatic masses obtained using our synthetic X-ray analysis outlined in Section~\ref{sec:syntheticxray} and their corresponding $r_{500,hse}$ to compute the SZ signal. 

The SZ signal is characterised by the value of its spherically integrated Compton parameter $d_{A}^{2}Y_{500}=(\sigma_{T}/m_{e}c^{2})\int PdV$ where $d_{A}$ is the angular diameter distance, $\sigma_{T}$ the Thomson cross-section, $c$ the speed of light, $m_{e}$ the electron rest mass, $P=n_{e}k_{B}T_{e}$ the electron pressure and the integration is done over the sphere of radius $r_{500}$.

All the simulations, produce fairly similar $Y_{500,hse}-M_{500,hse}$ relations, which are in reasonable agreement with the observations by \planck~of low-redshift massive clusters, in agreement with the results of \citet{Battaglia2012} and \citet{Kay2012}. Yet, as was the case for $Y_{X}$ in Section~\ref{sec:mass_Yx}, the large dynamic range in total SZ flux in the left panel of Fig.~\ref{fig:mass_Y500} gives a somewhat misleading impression of the sensitivity of the SZ signal to galaxy formation physics. Therefore, in the right panel of Fig.~\ref{fig:mass_Y500}, we normalize the total SZ signal by the self-similar expectation $\sigma_T/(m_e c^2 \mu_e m_p) f_b M_{500,hse} k_B T_{500,hse}$ (where $k_{B}T_{500,hse}\equiv\mu m_{p} G M_{500,hse} / 2 r_{500,hse}$) in order to remove the explicit gravitational halo mass dependence and to make more apparent any potential effects of baryonic physics upon the SZ signal--mass relation. The right panel of Fig.~\ref{fig:mass_Y500} clearly shows that the integrated SZ signal is sensitive to ICM physics.

In the \planck~cosmology, the standard AGN model reproduces the observed relation best of any of the radiative models (the unphysical \nocool~model performs similarly well, due to a conspiracy of having too high density and too low temperature). The scatter in the relation (which for clarity is only shown for the \agn~8.0 model) is also roughly reproduced. Thus, there is excellent consistency between the X-ray and SZ observables in terms of the physical story they tell.

It is worth noting that the sensitivity to baryonic physics increases with decreasing mass. We are currently conducting a detailed comparison to the stacked SZ signal--halo mass relation obtained by the Planck collaboration using $\sim 260,000$ Locally Brightest Galaxies taken from SDSS \citeauthor{Planck2012e}. As such a comparison requires synthetic SZ observations, its results will be presented elsewhere (Le Brun et al.\ in preparation).

%%%%%%%%%%%%%%%%%%%
% Optical and black hole properties %
%%%%%%%%%%%%%%%%%%%

\section{Optical and black hole scalings}

Finally, we compare the optical and black hole properties of the simulated systems to observations of local ($z\sim0$) groups and clusters. In Section~\ref{sec:fstar}, we look at the global stellar properties, then in Section~\ref{sec:BCG}, we investigate the optical properties of the brightest cluster galaxy (BCG) and, lastly, in Section~\ref{sec:BH}, we examine the properties of the central supermassive black hole.

%% Global stellar properties

\subsection{Total mass-to-light ratio}
\label{sec:fstar}

\begin{figure}
\begin{center}
\includegraphics[width=1.0\hsize]{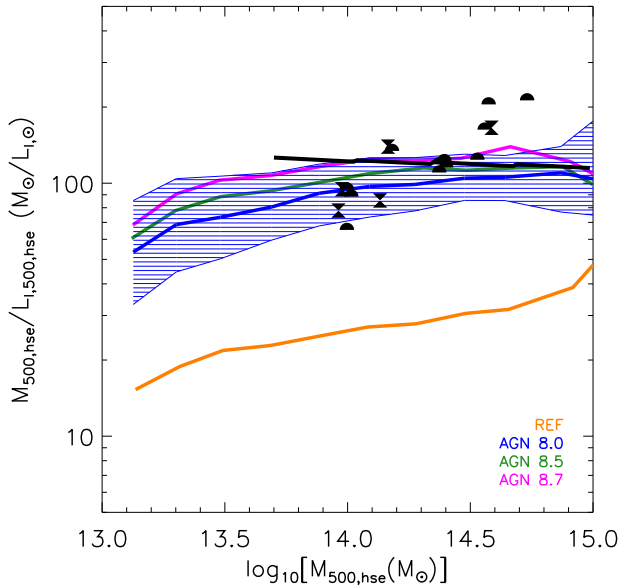}
\caption{\textit{I}-band total mass-to-light ratio as a function of $M_{500,hse}$ at $z=0$. The filled black hourglass and semi-circles represent the observational data of \citet{Sanderson2013} and \citet{Gonzalez2013}, respectively. The solid black line represents the SDSS image stacking results of \citet{Budzynski2014}. The three observational studies and the simulations include the contribution from intracluster light. The coloured solid curves represent the median \textit{I}-band total mass-to-light ratio--$M_{500,hse}$ relations in bins of $M_{500,hse}$ for the different simulations and the blue shaded region encloses 68 per cent of the simulated systems for the \agn~8.0 model. The observational studies differ in their findings of the {\it steepness} of the trend with halo mass, but consistently find high mass-to-light ratios for massive clusters. The inclusion of AGN feedback is essential for reproducing the observed high normalization.}
\label{fig:mass_fstar}
\end{center}
\end{figure}

In Fig.~\ref{fig:mass_fstar}, we plot the \textit{I}-band total mass-to-light ratio (within $r_{500,hse}$)--$M_{500,hse}$ relation at $z=0$ for the various simulations and compare to recent observations that explicitly include an intracluster light (ICL) component (we therefore avoid the difficulty of having to define what is the ICL in the simulations). To make like-with-like comparisons to the observations, we have computed Cousins \textit{I}-band luminosities using the \textsc{galaxev} model of \citet{Bruzual2003} (as described in Section~\ref{sec:OptObs}). As the observational total mass measurements of Fig.~\ref{fig:mass_fstar} are based on a hydrostatic analysis of X-ray data, we used the halo masses derived from our synthetic X-ray analysis. For the \citet{Gonzalez2013} and \citet{Sanderson2013} data (note that the Sanderson et al.\ sample is a subset of the Gonzalez et al.\ sample and uses their optical data, but the X-ray masses are computed somewhat differently), we have converted their stellar masses back into \textit{I}-band luminosities using their adopted stellar mass-to-light ratios. For the best-fit trend of \citet{Budzynski2014} (from their image stacking analysis), we use their derived \textit{I}-band stellar mass-to-light ratios (see their table 2) to convert their mean stellar masses into mean \textit{I}-band luminosities. We note that comparing luminosities should be more robust than comparing stellar masses, since stellar mass estimates rely on either dynamical mass-to-light ratios or stellar population synthesis modelling, which must assume a particular star formation history and metallicity (both must assume something about the stellar IMF as well). Both methods have significant ($\ga 0.1$ dex) systematic uncertainties. While going from stellar masses to luminosities in the simulations also requires a stellar population model, at least in this case we know the precise star formation history and metallicity of the star particles that make up the simulated galaxies, whereas these must be {\it assumed} for real galaxies. 

Observed galaxy clusters have high total mass-to-light ratios of $\sim 100$. Only the simulations that include feedback from supermassive black holes yield such high values. The \refsim~model, which neglects AGN feedback, produces mass-to-light ratios that are approximately a factor of three to five too low due to overly efficient star formation. These conclusions are insensitive to our choice of cosmology.

As discussed in detail in \citet{Budzynski2014}, there is a difference in the slope of the trend of the stellar mass/light with halo mass that they measure and that measured by \citet{Gonzalez2007} (and now \citealt{Gonzalez2013}). The origin of this difference is unclear. As noted by \citet{Budzynski2014} (see also \citealt{Leauthaud2012}), it is {\it not} driven by differences in the derived contributions of the ICL. Indeed, the largest differences are at the highest masses, where Gonzalez et al.~estimate that the ICL contributes a relatively small fraction of the total light. \citet{Budzynski2014} conclude that Gonzalez et al.\ consistently measure {\it lower} luminosities (and therefore higher total mass-to-light ratios) for the highest-mass systems compared to all the other observational studies they compared to (including \citealt{Lin2004a} and \citealt{Leauthaud2012}). Irrespective of this discrepancy, the observations strongly point to a high total mass-to-light ratio that cannot be achieved by means of stellar feedback alone.

%% BCG properties

\subsection{Properties of the BCGs}
\label{sec:BCG}

\subsubsection{Dominance of the BCG}

\begin{figure}
\begin{center}
\includegraphics[width=1.0\hsize]{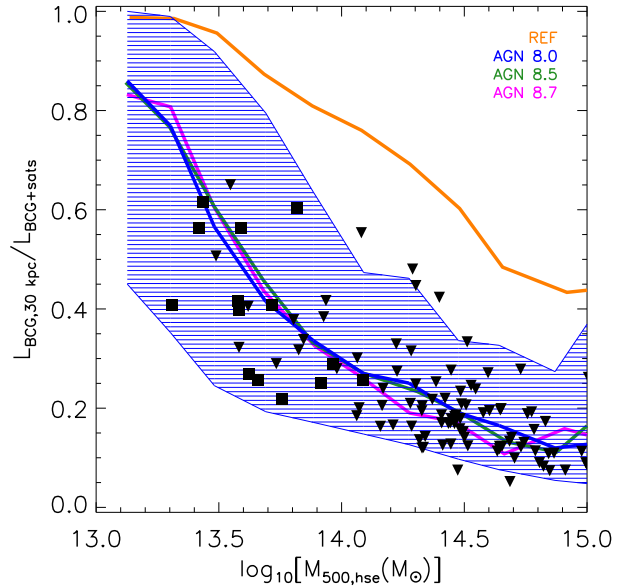}
\caption{\textit{K}-band luminosity fraction in the BCG at $z=0$. The filled black squares (groups) and downward triangles (clusters) represent the observational data of \citet{Rasmussen2009} and \citetalias{Lin2004b}, respectively. The coloured solid curves represent the median \textit{K}-band light fraction in the BCG--$M_{500,hse}$ relations in bins of $M_{500,hse}$ for the different simulations and the blue shaded region encloses 68 per cent of the simulated systems for the \agn~8.0 model. Lack of AGN feedback leads to BCGs which are too dominant compared to the satellite galaxy population.}
\label{fig:mass_fstarBCG}
\end{center}
\end{figure}

In Fig.~\ref{fig:mass_fstarBCG}, we plot the ratio of the \textit{K}-band light in the BCG to that in the BCG and satellite galaxies (i.e. no ICL) as a function of halo mass at $z=0$ for the various simulations and compare to observations of individual X-ray-selected systems of \citet{Lin2004a} and \citet*{Lin2004b} (hereafter collectively referred to as \citetalias{Lin2004a}) and \citet{Rasmussen2009}. In both cases, we have converted the observed mean X-ray temperature into a halo mass using the mass--temperature relation of \citet{Vikhlinin2009}. For the simulations, we compute the \textit{K}-band light of the BCG in a simple way by summing the luminosities of all the star particles within an aperture of 30 kpc. This is similar to the average effective radius of observed BCGs (e.g. \citealt{Stott2011}). Adjusting the aperture changes the normalization of the relation somewhat but does not significantly affect the shape of the relation. 

Note that in both the observations and simulations plotted in Fig.~\ref{fig:mass_fstarBCG}, the BCG is defined to be the most (stellar) massive/luminous galaxy, and that there is no requirement that the BCG be, for instance, coincident with the X-ray emission peak or the `central' galaxy. Indeed, recent observational studies (e.g. \citealt{Skibba2011,Balogh2011}) have shown that there can sometimes be relatively large offsets between the BCG and these other choices of centre . 

As can clearly be seen, the stellar fraction in the BCG is a strongly decreasing function of total mass. All the models reproduce that trend, but the \refsim~model produces BCGs which are too dominant compared to the observed ones, whereas all the \agn~models yield similar stellar fractions in the BCGs which are consistent with the observed ones. This is due both to suppression of star formation in massive satellite galaxies which eventually merge with the BCG, as well as to the suppression of the central cooling flows by the AGN feedback. As we will show in the next subsection, central cooling flows and the star formation they induce in BCGs are indeed strongly suppressed by AGN feedback. No reasonable choice of aperture can reconcile the observed trend with the \refsim~model.

\subsubsection{star-forming fraction }

\begin{figure}
\begin{center}
\includegraphics[width=1.0\hsize]{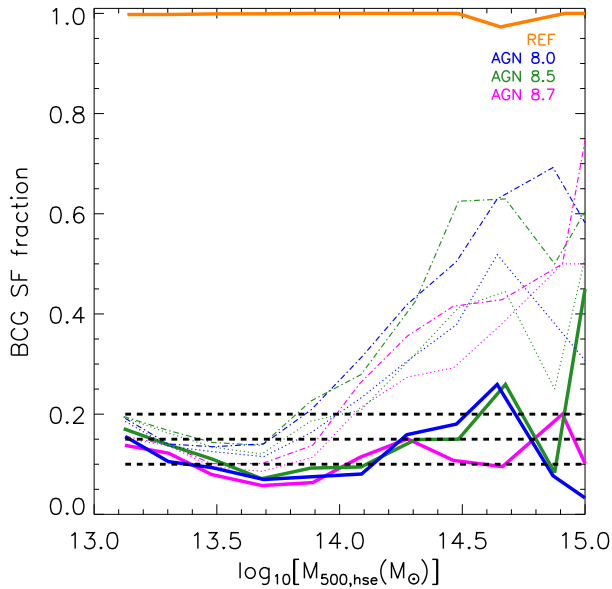}
\caption{Fraction of the BCGs that are currently forming stars at an appreciable rate ($SFR>3~\textrm{M}_{\odot}\textrm{ yr}^{-1}$) as a function of $M_{500,hse}$. The black dashed lines correspond to the observational results of \citet{Edwards2007}. The thick solid, thin dotted and dot-dashed curves (orange, blue, green and magenta) represent the median relations for the different simulations in 10, 20 and 30 kpc apertures, respectively. The observed star-forming fraction  is roughly reproduced in the AGN models when measured approximately within an observed aperture. However, the star-forming fraction  increases with halo mass when the aperture is enlarged.}
\label{fig:bcgsffraction}
\end{center}
\end{figure}

In Fig.~\ref{fig:bcgsffraction}, we plot the fraction of the BCGs that are currently forming stars at an appreciable rate ($SFR>3~\textrm{M}_{\odot} \textrm{ yr}^{-1}$) as a function of system mass for the various simulations. We compare to the observations of the BCGs of both X-ray-selected groups and clusters \citep[from the National Optical Astronomy Observatory Fundamental Plane Survey (NFPS);][]{Smith2004} and optically-selected groups and clusters from the SDSS Data Release 3 (DR3) C4 cluster catalogue \citep{Miller2005} by \citet{Edwards2007} (black dashed lines). The thick solid, thin dotted and dot-dashed curves represent the median relations for the simulations when respectively a 10, 20 and 30~kpc aperture is used to define the BCG.

\citet{Edwards2007} find that the star-forming fraction  of BCGs (i.e. those with detectable optical line emission, corresponding to a SFR threshold of a few solar masses per year) is approximately independent of system mass. The spectroscopic measurements are made within 2 or 3 arcsecond fibres, which at the typical redshifts of the NFPS and C4 samples corresponds to an aperture of a few kpc across. When we compute the star-forming fraction  in a similar aperture (solid thick curve in Fig.~\ref{fig:bcgsffraction}), we find a similar trend and normalization to the observed one for the models that include AGN feedback. The \refsim~model, which only includes stellar feedback, fails to suppress the central cooling flows and their induced star formation in BCGs. However, as demonstrated by the dotted and dot-dashed curves, when the aperture is expanded, the star-forming fraction  begins to rise with halo mass. Although we are unaware of any observations that show that such large-scale star formation does not exist in general in real BCGs, we suspect this trend may be at least partly numerical in origin. Specifically, we have examined the maximum past temperature (the simulation code tracks this quantity for each particle over all time steps) of star-forming gas and recently-formed star particles (those formed within the past Gyr) within the annulus 10 kpc $< r \le$ 30 kpc centred on the BCG. The vast majority of the particles have a maximum past temperature of just below $10^{5.0}$ K, corresponding to the temperature floor imposed by UV/X-ray photoheating in the simulations, with a further contribution from gas with a maximum past temperature between $10^{5.5}$ K and $10^{6.0}$ K and a negligible contribution from gas with a maximum past temperature between $10^{6.5}$ K and $10^{7.5}$ K. In short, the recent extended star formation is being driven by gas that was {\it never} part of the hot ICM. Instead, the gas was stripped by orbiting satellites (e.g. \citealt{Puchwein2010}) and the reason why more massive clusters are more likely to have extended star formation is simply because there are more satellites to deposit cold gas in this fashion.  However, this extended star formation may be numerical in origin, as it is known that standard SPH inherently suppresses mixing through, for instance, the Kelvin-Helmholtz instability (e.g. \citealt{Agertz2007,Mitchell2009}), which might otherwise dissolve the cold gas clumps.

%Although we are unaware of any observations that show that such large-scale star formation does not exist in real BCGs, we suspect that this trend indicates that the f in the simulations are not completely effective at preventing thermal instabilities from developing (presumably because the heating is not spatially extended enough). Alternatively, the cold, star-forming gas  may have been deposited at large radii from stripped satellite galaxies (e.g. \citealt{Puchwein2010}).

\subsubsection{Colour}

\begin{figure}
\begin{center}
\includegraphics[width=1.0\hsize]{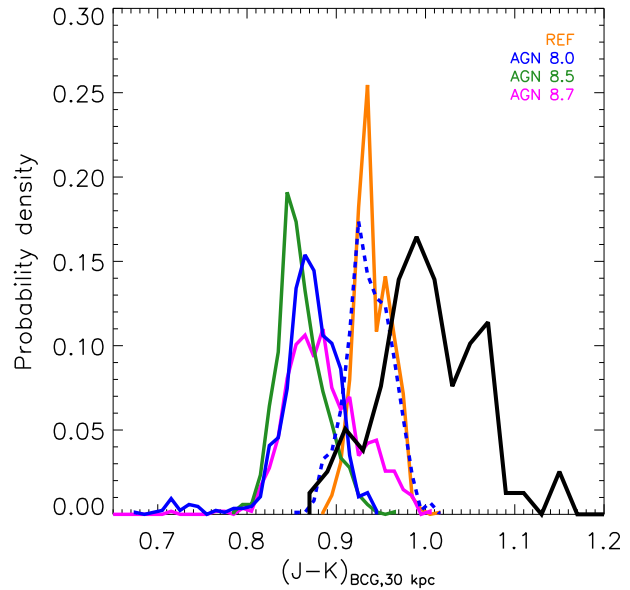}
\caption{Distribution of BCG rest-frame $J-K$ colour at $z=0$. The thick solid histograms (orange, blue, green and magenta) are for the different simulations while the black one corresponds to the observational data of \citet{Stott2008} with $z\le0.25$. The blue dashed histogram corresponds to the \agn~8.0 model when the metallicity of each of the star particles is doubled. All models produce BCGs with $J-K$ colours that are too blue, signalling that the empirical nucleosyntethic yields and/or the SNIa rates adopted in the simulations may be somewhat too low.}
\label{fig:colour}
\end{center}
\end{figure}

In Fig.~\ref{fig:colour}, we plot the $z=0$ distribution of the $J-K$ BCG colours for the various simulations and compare to the observations from the X-ray-selected rich galaxy clusters of \citet{Stott2008}. The $J-K$ colours of \citet{Stott2008} are observer-frame colours. In order to reliably compare with the simulation rest-frame colours at $z=0$, we have K-corrected them to the $z=0$ rest-frame using the {\textsc{calc\_kcor}} \textsc{idl} routine, which is based upon the analytical approximation of \citet*{Chilingarian2010} and \citet{Chilingarian2012}. In addition, as \citet{Stott2008} have selected BCGs whose host clusters have $L_X>10^{44}~\textrm{erg s}^{-1}$, we use only the BCGs with $M_{500,hse}\ge 10^{14}~\textrm{M}_{\odot}$ (which roughly corresponds to $L_X=10^{44}~\textrm{erg s}^{-1}$ according to Fig.~\ref{fig:mass_Lx}). The benefit of using $J-K$ is that it is relatively insensitive to dust attenuation as well as to `frosting' due to recent low levels of star formation (since it is probing mainly old main sequence stars).

\begin{figure*}
\begin{center}
\includegraphics[width=0.49\hsize]{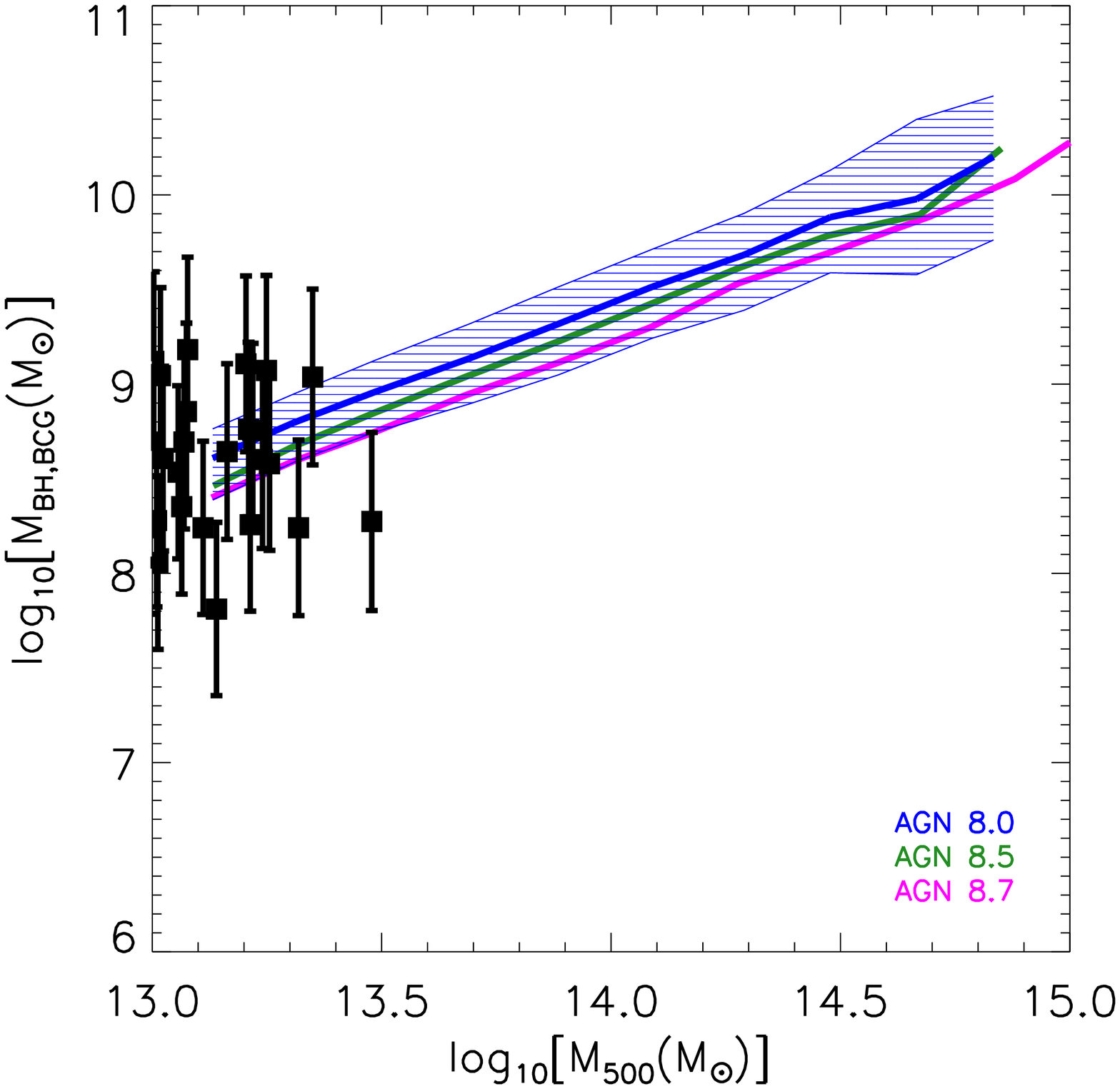}
\includegraphics[width=0.49\hsize]{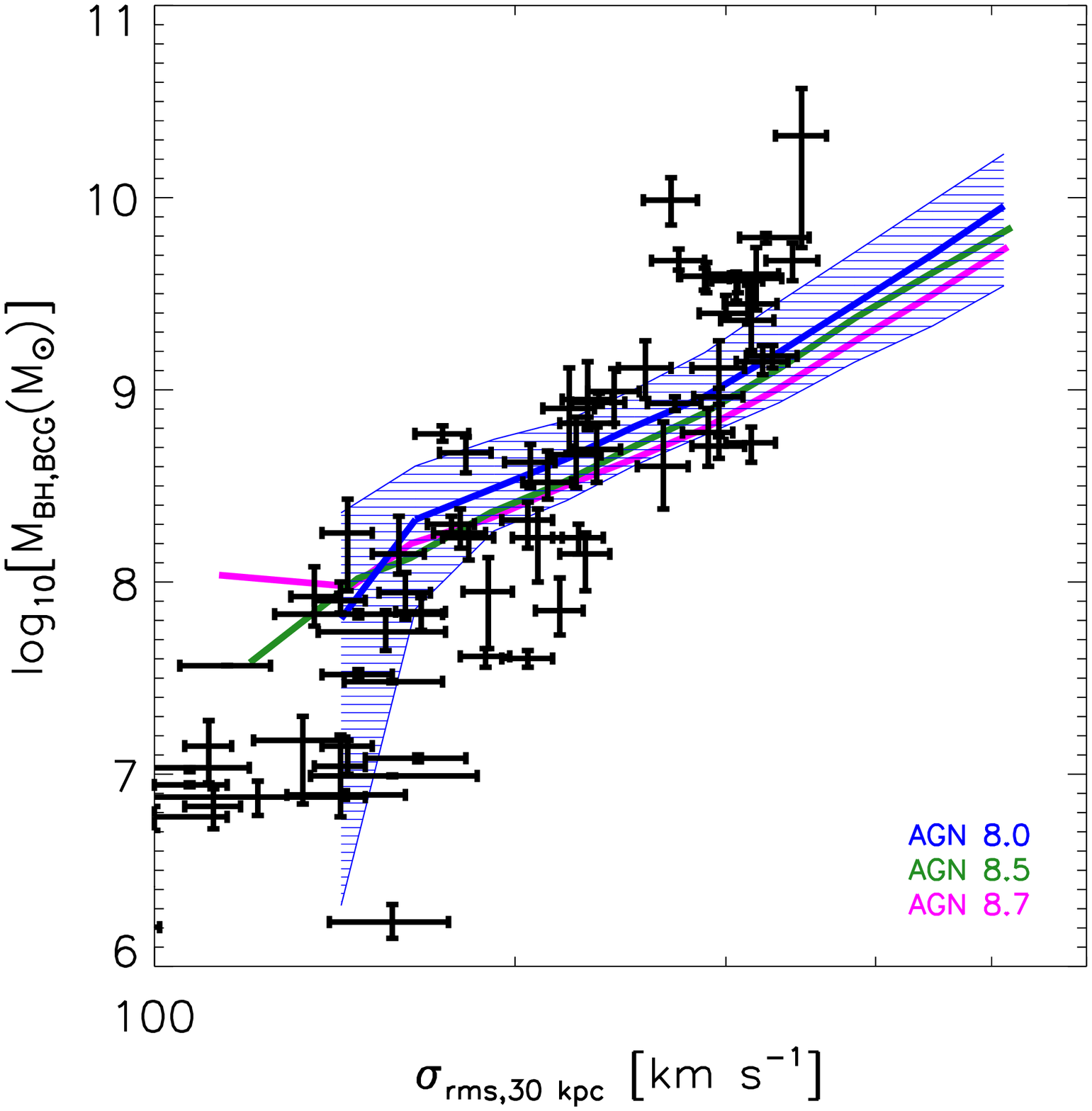}
\caption{Mass of the central supermassive black hole as a function of $M_{500}$ (\emph{left}) and of the root mean square one-dimensional stellar velocity dispersion of the BCG in a 30 kpc aperture (\emph{right}). The filled black circles and squares with error bars correspond to the observational data of \citet{McConnell2013} and \citet{Bandara2009}, respectively. The coloured solid curves represent the median central supermassive black hole mass in bins of $M_{500}$ or stellar velocity dispersion for the different simulations and the blue shaded region encloses 68 per cent of the simulated systems for the \agn~8.0 model. The AGN models broadly reproduce the normalization of the observed black hole scaling relations.}
\label{fig:bh}
\end{center}
\end{figure*}

Surprisingly, all the models produce BCGs with $J-K$ colours that are too blue compared to the observations of \citet{Stott2008}, by about 0.15 dex on average in the case of the AGN models. One possible reason for this discrepancy is that the simulated BCGs may have unrealistically low metallicities. Indeed, \citet{McCarthy2010} found that the central galaxies of simulated groups in OWLS had too low metallicity by about 0.5 dex (we confirm that this holds true here as well). As discussed by McCarthy et al., this could plausibly be explained by the adoption of nucleosynthetic yields and/or SNIa rates in the simulations that are too low. Both were chosen based on empirical constraints but have uncertainties at the factor of 2 level each \citep{Wiersma2009b}. We have therefore tried boosting the metallicity of the star particles by factors of two and three (in post-processing for the \agn~8.0 model when computing the $J-K$ colours. (Note it is the high metallicity of the BCGs in the \refsim~model which makes them somewhat redder than the BCGs in the AGN models, in spite of their higher star formation rates -- but as shown by \citealt{McCarthy2010}, the \refsim~model BCGs have too {\it high} metallicities compared to observations.) This indeed reduces the level of disagreement: when the stellar metallicities are doubled (blue dashed line), the colours are too blue by $\approx 0.075$ dex on average (i.e. the level of disagreement is halved); while when they are tripled, the peaks of the observed and simulated distributions are roughly in the same position (i.e. the average discrepancy has nearly disappeared), but the observed distribution has a larger tail towards red colours. 

It is unclear what the origin of the remaining discrepancy (the tail towards redder colours) is. We have experimented with a variety of stellar population synthesis models using the online tool EzGal\footnote{http://www.baryons.org/ezgal/}. Conservatively adopting simple stellar populations, we are unable to produce rest-frame $J-K$ colours $\ga 1.0$ for even fairly extreme choices of the formation redshift (e.g. $z_f = 5$) and super-solar metallicities ($Z = 1.5~\textrm{Z}_\odot$). This suggests that either there is a systematic error inherent to current stellar population synthesis models and/or there is an issue with the observed colours. One possible cause of redder colours could be relatively large amounts of dust either in the BCG itself or along the line of sight, which have not been accounted for.

It is worth noting that we are computing the colours using 30~kpc apertures, which contain extended star formation (see Fig.~\ref{fig:bcgsffraction}). We have checked that reducing the aperture size (to both 10 and 20~kpc) cannot explain the discrepancy. It only shifts the maximum of the distribution by no more than $\sim0.03$ dex and does not seem to affect the position of its peak (note that $J-K$ is generally insensitive to recent star formation).

%% Black hole scalings

\subsection{Black hole scalings}
\label{sec:BH}

In the left panel of Fig.~\ref{fig:bh}, we plot the relation between the mass of the BCG central supermassive BH and $M_{500}$ for the various simulations which include AGN feedback and compare to the observations of individual strong gravitational lenses of \citet*{Bandara2009}. As their mass measurements (we have converted their $M_{200}$ masses into $M_{500}$ assuming a NFW profile with a concentration of $4$, which yields $M_{500} \approx 0.69 M_{200}$) come from strong lensing, we use the true $M_{500}$ for the simulated systems for this comparison. In the right panel of Fig.~\ref{fig:bh}, we plot the mass of the BCG's central supermassive black hole as a function of the one-dimensional BCG velocity dispersion in a 30 kpc aperture for the various simulations and compare to the recent compilation of the properties of 72 central black holes and their host galaxies of \citet{McConnell2013}.

Both relations are reasonably reproduced by the three AGN feedback models considered here. The fact that the normalizations of the BH scaling relations are well reproduced is not too surprising, as the efficiency of the feedback $\epsilon$ was tuned by \citet{Booth2009} roughly to match the normalization of $m_{BH}-m_{halo}$ relation at $z=0$ as well as the present-time cosmic BH density (see also Appendix~\ref{sec:reso}). They also showed that the simulations roughly reproduce the normalization of the $z=0$ relations between BH mass, stellar mass and velocity dispersion. It was nevertheless worth checking that the calibration which was done using smaller simulations (up to $100~h^{-1}$ Mpc) with higher mass resolution (up to 8 times higher) remains valid for simulations with larger volume and lower mass resolution. This shows that supermassive BHs are still able to regulate their growth even though the simulation volume has been increased and the mass resolution decreased. Finally, the fact that the three AGN models yield similar scaling relations means that we have not increased the heating temperature beyond the value at which the supermassive BHs are no longer able to regulate their growth, because the time between heating events exceeds the Salpeter time-scale for Eddington-limited accretion \citep[see][]{Booth2009}.

We note that there is an hint of a difference in the slopes of the observed and simulated trends in the right panel of Fig.~\ref{fig:bh}. It is unclear whether this difference is real or not, as we have not mimicked a full observational selection and analysis of the simulated systems. Furthermore, the observed velocity dispersion is generally measured on smaller scales (e.g. inside the galaxy's half-light radius) than can be reliably done with the current simulations, due to their limited resolution.

\section{Summary and Discussion}

We have presented a new suite of large volume ($400~h^{-1}$ Mpc on a side) cosmological hydrodynamical simulations (called cosmo-OWLS, an extension to the OverWhelmingly Large Simulations project; \citealt{Schaye2010}) which has been specifically designed to aid our understanding of galaxy cluster astrophysics and thereby attempt to minimize the main systematic error in using clusters as probes of cosmology. We have investigated five different physical models: a non-radiative model (\nocool), a model which includes metal-dependent radiative cooling, star formation and stellar feedback (\refsim) and three models which further include AGN feedback with increasing heating temperatures (from \agn~8.0 with $\Delta T_{heat}=10^{8}~\textrm{K}$ to \agn~8.7 with $\Delta T_{heat}=10^{8.7}~\textrm{K}$ through \agn~8.5 with $\Delta T_{heat}=10^{8.5}~\textrm{K}$).

In this first paper, we have made detailed comparisons to the observed X-ray, Sunyaev--Zel'dovich effect, optical, and central supermassive black hole properties of local groups and clusters. In order to make like-with-like comparisons, we have produced synthetic observations and mimicked observational analysis techniques. For instance, we have not only computed X-ray spectra for each of the simulated systems and fitted single-temperature plasma models to them in order to obtain metallicity, temperature and density profiles, but also conducted a hydrostatic mass analysis using the best-fitting temperature and density profiles and the functional forms of \citet{Vikhlinin2006}. From these comparisons, we conclude the following: 
\begin{enumerate}
\item AGN feedback is essential for reproducing the strong trend in the observed gas fractions with halo mass (Fig.~\ref{fig:mass_fgas}) and the high total mass-to-light ratios (i.e. low star formation efficiencies) of groups and clusters (Fig.~\ref{fig:mass_fstar}). All of our models consistently predict a weak dependence of the star formation efficiency on halo mass, in accordance with the trends observed by \citet{Budzynski2014} (see also \citealt{Leauthaud2012}) but significantly shallower than the trend derived by \citet{Gonzalez2013}.
\item In the \planck~cosmology, the fiducial AGN model (\agn~8.0) reproduces the global hot gas properties over approximately two orders of magnitude in halo mass ($10^{13}~\textrm{M}_\odot \la M_{500} \la 10^{15}~\textrm{M}_\odot$), including the observed luminosity--mass, mass--temperature, $f_{\rm gas}$--mass, $Y_X$--mass, and SZ flux--mass trends (Figs.~\ref{fig:mass_Lx} to \ref{fig:mass_Yx} and \ref{fig:mass_Y500}, respectively). For the first time, the simulations also broadly reproduce the observed scatter. Higher AGN heating temperatures (leading to more violent, bursty feedback when using the OWLS implementation of AGN feedback) lead to under-luminous (and slightly overheated) and under-dense clusters with lower-than-observed SZ fluxes, although this can be mitigated to an extent by appealing to a higher universal baryon fraction (e.g. as in the \wmap7 cosmology).
\item Contrary to previous claims, we find that the SZ flux (Fig.~\ref{fig:mass_Y500}) and its X-ray analogue $Y_X$ (Fig.~\ref{fig:mass_Yx}) {\it are} sensitive to baryonic physics. In particular, gas ejection by AGN can significantly reduce both quantities (the corresponding increase in temperature resulting from the ejection of low-entropy gas is not sufficient to compensate for the lower gas density if large quantities of gas are ejected). This serves as a warning against blindly applying $Y_X$, SZ flux, and gas mass (fraction) scalings to low halo masses ($M_{500} \la 10^{14}~\textrm{M}_\odot$) and/or high redshifts without an independent mass check.
\item The fiducial AGN model reproduces not only the global hot gas properties over two decades in mass, but also the observed density and entropy (and therefore also temperature and pressure) radial distributions of the ICM over 1.5 decades in radius, from $0.05 \la r/r_{500} \la 1.5$, over this mass range (Figs.~\ref{fig:Sprof} to \ref{fig:rhoprof}). To our knowledge, this is the first time a cosmological hydrodynamical simulation has reproduced the detailed radial distribution of the hot gas, including the central regions. 
\item The fiducial AGN model also reproduces the observed large scatter in the central density distribution of the hot gas. Interestingly, the central gas density shows no evidence for significant bimodality (Fig.~\ref{fig:n0}).
\item AGN feedback is essential not only to lower the overall star formation efficiencies of groups and clusters, but also to reduce the dominance of the brightest cluster galaxy (BCG) with respect to the satellite population, and to prevent significant present-day star formation (Figs.~\ref{fig:mass_fstar} to \ref{fig:bcgsffraction}).
\item While successfully shutting off cooling in the very central regions of the BCG in accordance with observations, the simulated BCGs have low levels of spatially-extended star formation (Fig.~\ref{fig:bcgsffraction}), which is being driven by recently deposited cold gas (ISM) from ram pressure-stripped satellite galaxies.  This trend may be at least partly numerical in origin, due to suppression of mixing (e.g. via the Kelvin-Helmholtz instability) in standard SPH.
\item The simulated BCGs, while having approximately the correct stellar mass and central star-forming fraction, are too blue in $J-K$ (by about 0.15 dex on average; Fig.~\ref{fig:colour}) compared to observed local BCGs \citep{Stott2008}. This discrepancy may be due to adopting incorrect yields and/or SNIa rates in the simulations (which are based on empirical constraints that have uncertainties at the factor of $\approx 2$ level). Tripling the stellar metallicities for the \agn~8.0 model brings the position of the peak of the distribution into agreement with the peak of the observed distribution. 
\item The simulations broadly reproduce the observed black hole mass -- halo mass -- velocity dispersion relations (Fig.~\ref{fig:bh}). The feedback efficiency was calibrated by \citet{Booth2009} to approximately match the normalization of these relations in higher resolution simulations and at lower halo masses. Here we show that the agreement continues to hold at much larger masses and somewhat lower resolution. Neither the black hole feedback efficiency nor the accretion model were tuned in any way to reproduce the properties of galaxy groups and clusters.
\end{enumerate}

The success of the fiducial AGN model in reproducing the detailed hot gas properties over 1.5 decades in radius and the global hot gas and global optical properties over two decades in halo mass, as well as the system-to-system scatter in the X-ray/SZ properties, is an important step forward. The production of reasonably realistic simulated {\it populations}, as well as models that bracket the observations, opens the door to producing synthetic cluster surveys to aid the astrophysical and cosmological interpretation of up-coming/on-going cluster surveys and to help quantify the important effects of observational selection. We are using cosmo-OWLS for precisely this purpose and intend to make synthetic X-ray, SZ, optical, and lensing surveys available in the near future.

The predicted hot gas and stellar properties are highly model dependent. Indeed, even for a fixed sub-grid AGN feedback efficiency, i.e. for models that inject a fixed amount of energy per unit of accreted gas mass, the effective efficiency of the AGN feedback is sensitive to the way in which the energy is injected. A higher heating temperature, which corresponds to less frequent but more energetic outbursts, results in more efficient feedback.  As dicussed in Section 2.1, we anticipated that using increased heating temperatures may be necessary to avoid overcooling in the most massive clusters, where $T_{\rm vir} \sim 10^8$ K.  However, increasing the heating temperature had a large effect on the {\it progenitors} of these (massive galaxies and low-mass groups at $z \sim 2$) which in turn had important knock-on effects for the $z=0$ population of massive clusters (most importantly significantly reduced gas fractions).  The complicated merger history of clusters makes it difficult to anticipate these results.  In any case, the demonstrated sensitivity to model parameters means that the models must continue to be challenged with new observables (e.g. detailed properties of the satellite galaxy population, which we have not explored here) and over a wider range of masses and redshifts than we have considered here. In addition, quantitative comparisons of the simulations to the observations (rather than the rough `by eye' evaluations presented here) require careful consideration of observational selection effects, particularly in the group regime. 

From the comparisons we have made thus far (both here and in \citealt{McCarthy2010,McCarthy2011}), the total mass-to-light ratio (star formation efficiency) appears to be the best discriminator for distinguishing between the impact of different sources of feedback (stellar feedback vs AGN). However, the detailed hot gas properties are more sensitive to the nature of the AGN feedback than are the stellar properties or BH scaling relations. In particular, given that the fiducial model reproduces the observations significantly better than models with higher heating temperatures, this suggests that the AGN feedback mechanism in real clusters is/was similarly violent and bursty as in this model. An independent test of the models will therefore be to compare to the demographics of the observed AGN population (e.g. `radio' vs `quasar' mode duty cycles and luminosity functions and their dependencies on redshift and environment).

In a companion paper \citep{McCarthy2014}, we examine the predictions of the cosmo-OWLS suite for the thermal Sunyaev--Zel'dovich effect power spectrum and make comparisons with recent measurements thereof.

\section*{Acknowledgements}

The authors would like to thank the members of the OWLS team for their contributions to the development of the simulation code used here and the anonymous referee for a constructive report. They also thank Ming Sun and Yen-Ting Lin for providing their observational data and August Evrard, Andrey Kravtsov, Marguerite Pierre, and Ramin Skibba for helpful discussions. AMCLB acknowledges support from an internally funded PhD studentship at the Astrophysics Research Institute of Liverpool John Moores University. IGM is supported by an STFC Advanced Fellowship at Liverpool John Moores University. JS is sponsored by the European Research Council under the European Union's Seventh Framework Programme (FP7/2007-2013)/ERC Grant agreement 278594-GasAroundGalaxies. 
This work used the DiRAC Data Centric system at Durham University, operated by the Institute for Computational Cosmology on behalf of the STFC DiRAC HPC Facility (www.dirac.ac.uk). This equipment was funded by BIS National E-infrastructure capital grant ST/K00042X/1, STFC capital grant ST/H008519/1, and STFC DiRAC Operations grant ST/K003267/1 and Durham University. DiRAC is part of the National E-Infrastructure.

\bibliographystyle{mn2e}
\bibliography{scalings}

\appendix
\section{Resolution study}
\label{sec:reso}

\begin{figure*}
\begin{center}
\includegraphics[width=0.49\hsize]{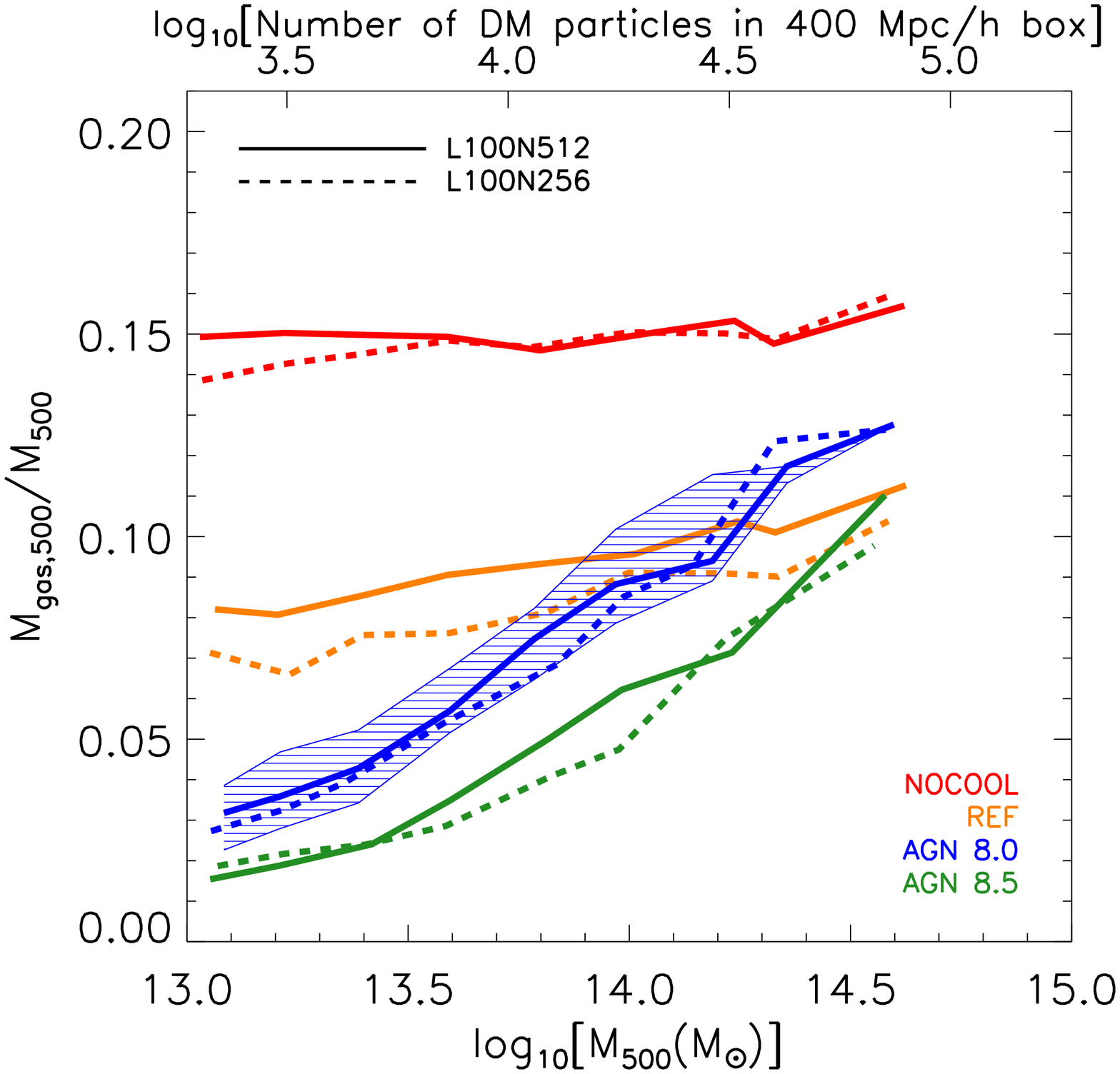}
\includegraphics[width=0.49\hsize]{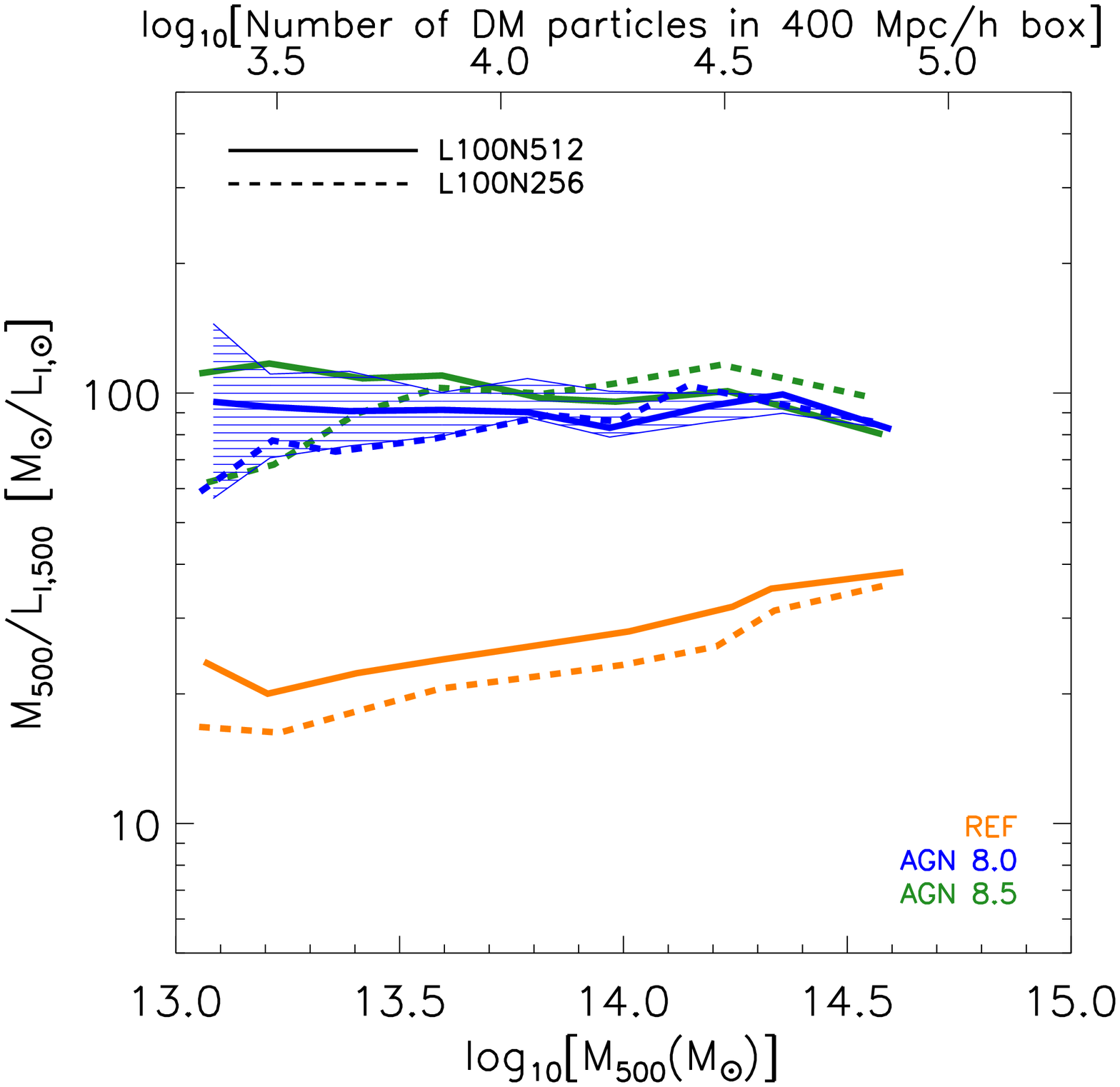}
\caption{Effect of numerical resolution on the median gas mass fraction--$M_{500}$ and $I$-band total mass-to-light ratio--$M_{500}$ relations at $z=0$. The simulations used here assume the \wmap7 cosmology. Global properties are adequately converged down to $\log_{10}[M_{500}(\textrm{M}_\odot)]\sim13.3$ (i.e. a few times $10^{13}~\textrm{M}_{\odot}$). Both panels use the true physical properties (gas fraction, total mass and $I$-band total mass-to-light ratio) of the simulated systems (i.e. no synthetic observations were used).}
\label{fig:reso}
\end{center}
\end{figure*}

\begin{figure}
\begin{center}
\includegraphics[width=1.0\hsize]{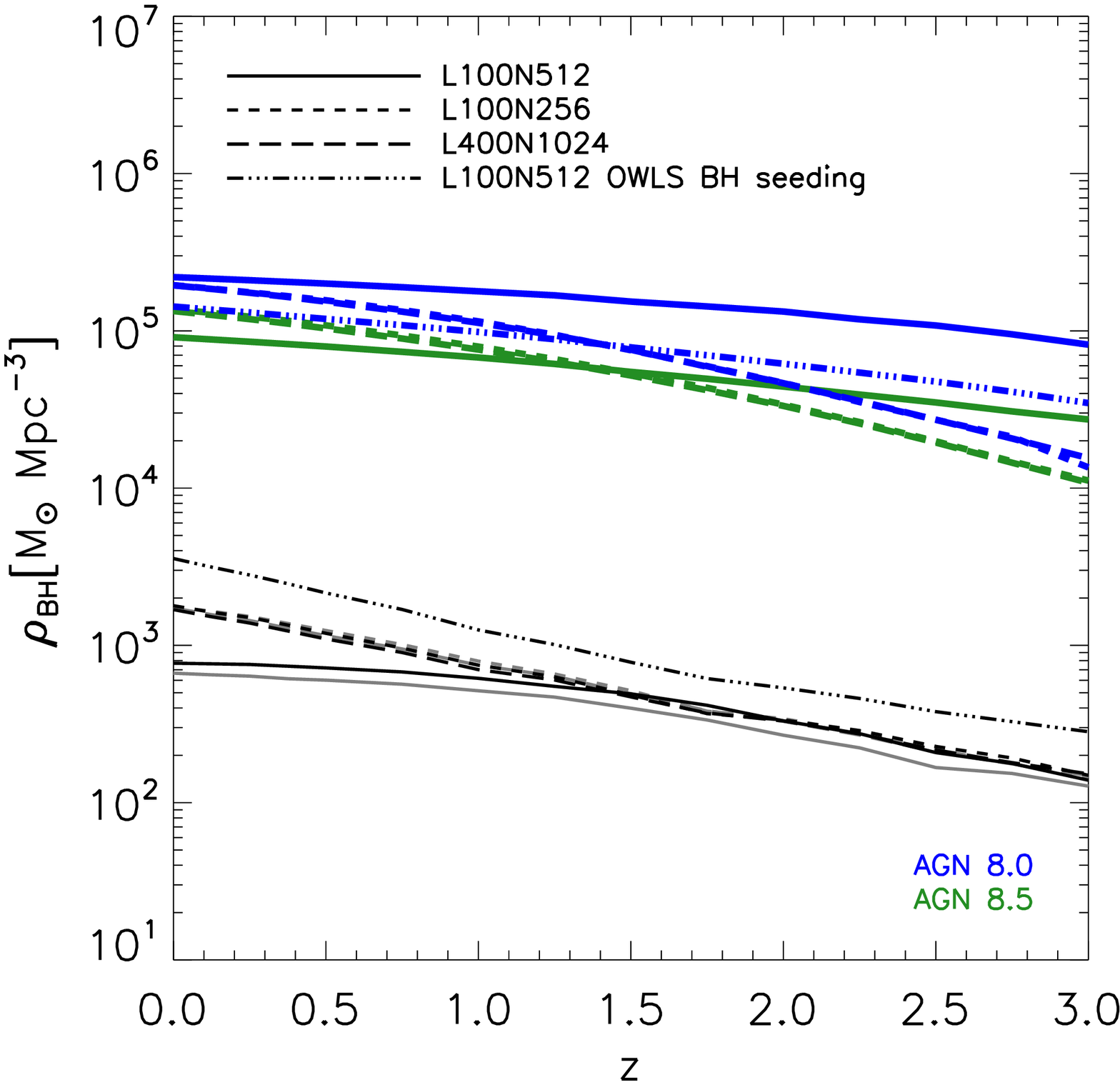}
\caption{Effect of box size, numerical resolution and BH seeding on the evolution of the cosmic BH density. The simulations used here assume the \wmap7 cosmology. The black and grey curves show the cumulative density in seed BHs for the \agn~8.0 and \agn~8.5 models, respectively. The solid lines correspond to the simulations run in a $100~h^{-1}$ Mpc on a side box at eight times higher mass resolution than the production runs. The dashed lines and long-dashed lines (which are virtually on top of each other) correspond to the simulations run in $100~h^{-1}$ Mpc and $400~h^{-1}$ Mpc on a side boxes at the resolution of the production runs, respectively. All these simulations use the same halo mass limit for BH particle and BH seed mass as the production runs. The triple-dot-dashed lines correspond to the high resolution runs but with the BHs injected in eight times less massive haloes and with a eight times lower seed mass as they were originally in the OWLS \agn~model (see \citet{Booth2009} and Section~\ref{sec:sims}).}
\label{fig:rhoBH}
\end{center}
\end{figure}

We examine the sensitivity of our results to numerical resolution. As currently available hardware prevents us from running higher resolution simulations in $400~h^{-1}$ Mpc on a side boxes, we use smaller simulations for testing numerical convergence. They are $100~h^{-1}$ Mpc on a side and use $2\times256^3$ particles (which is the same resolution as our $2\times1024^{3}$ particles in $400~h^{-1}$ Mpc box runs) and $2\times512^3$ particles (i.e. eight times higher mass resolution and two times higher spatial resolution). They assume the \wmap7 cosmology. Note that when comparing AGN models at different resolutions, we adopt the same {\it halo} mass limit for BH particle seeding and BH seed mass (see Section~\ref{sec:sims} for seeding details) and that the convergence tests are made using the true physical properties of the simulated systems (i.e. no synthetic observations were used).

In Fig.~\ref{fig:reso}, we compare the median gas mass fraction--$M_{500}$ (\emph{left}) and $I$-band total mass-to-light ratio--$M_{500}$ (\emph{right}) relations at $z=0$ for systems with $12.9 \la \log_{10}[M_{500}(\textrm{M}_\odot)] \la 14.75$ at the resolution of the production runs (dashed lines) and at eight times higher mass resolution (solid lines) for four of the models used (\nocool, \refsim, \agn~8.0 and \agn~8.5). We find that global properties are adequately converged down to $\log_{10}[M_{500}(\textrm{M}_\odot)]\sim13.3$ (i.e. a few times $10^{13}~\textrm{M}_{\odot}$) at the resolution of the cosmo-OWLS runs. 

In Fig.~\ref{fig:rhoBH}, we compare the evolution of the global BH density and of the cumulative BH density present in seed-mass BHs (black and grey curves) when box size, resolution and BH seeding are varied for both the \agn~8.0 and \agn~8.5 models. Varying box size at fixed resolution and seeding parameters from $100~h^{-1}$ Mpc (dashed lines) to $400~h^{-1}$ Mpc (long-dashed lines) on a side has no noticeable effect upon the evolution of the global BH density and cumulative density in seed BHs for $z\le3$ for both the \agn~8.0 and \agn~8.5 models (i.e., the dashed lines and long-dashed lines lie on top of each other). Varying resolution at fixed box size and seeding parameters from the resolution of the production runs (dashed lines) to eight times higher mass resolution (solid lines) affects the evolution of both densities up to the present time in both \agn~models. Finally, varying the halo mass limit for BH particle seeding and the BH seed mass from the values used for the original OWLS AGN model (triple-dot-dashed lines) to eight times higher masses as used for the production runs (solid, dashed and long-dashed lines) at fixed box size and mass resolution (solid lines), leads to higher BH and seed BH densities at all redshifts.  

Overall, however, the differences are not that large between the different models and all are approximately consistent with the observational constraints on the $z\approx0$ mass density of SMBHs of \citet{Shankar2004}.

\section{Hydrostatic bias and spectroscopic temperatures}
\label{sec:biases}

\begin{figure}
\begin{center}
\includegraphics[width=1.0\hsize]{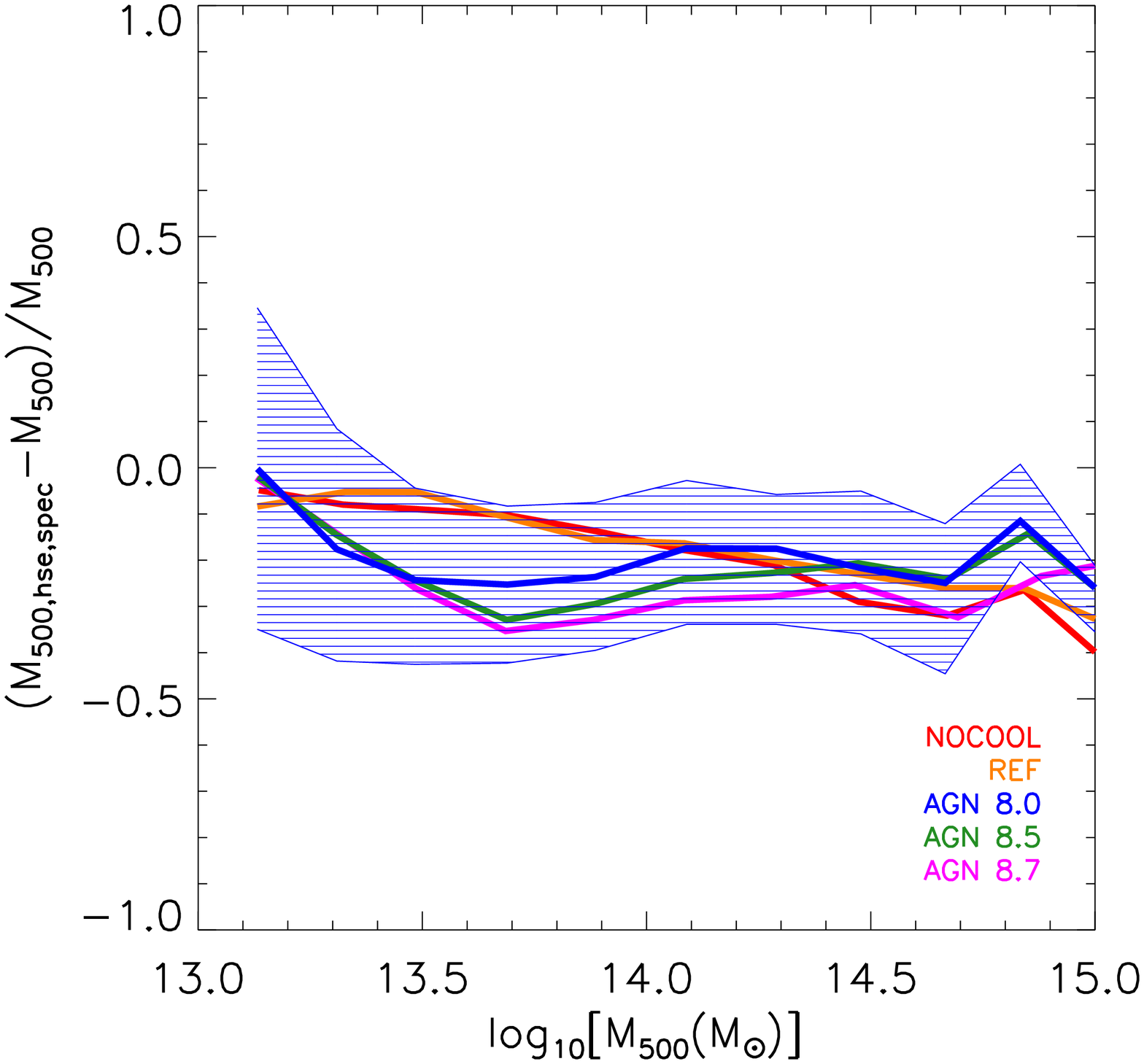}
\caption{Hydrostatic bias as a function of $M_{500}$ at $z=0$. Consistent with previous simulation studies, we find a mean bias of $\sim- 20$ per cent for both groups and clusters. The scatter increases with decreasing total mass.}
\label{fig:hsebias}
\end{center}
\end{figure}

\begin{figure}
\begin{center}
\includegraphics[width=1.0\hsize]{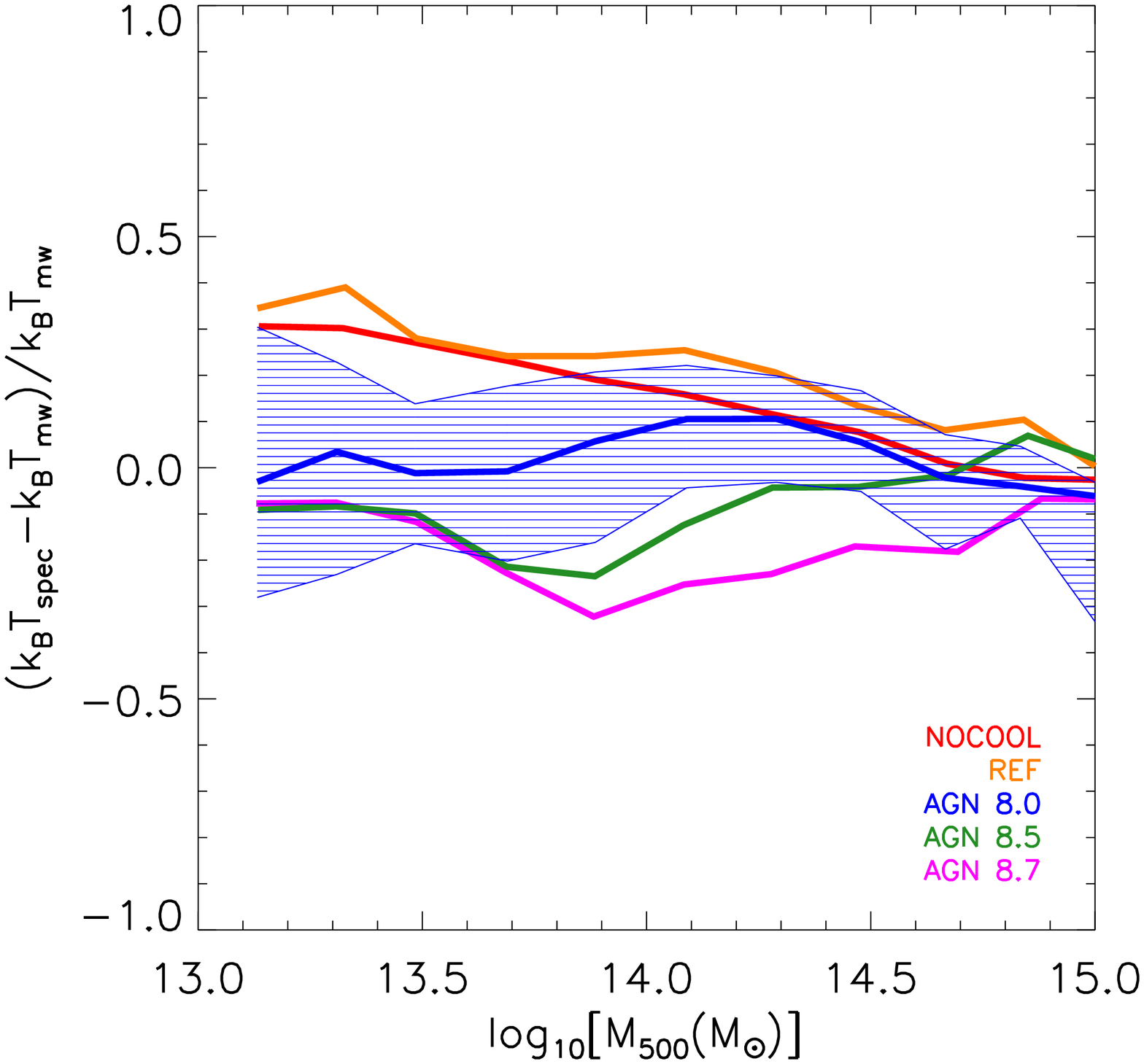}
\caption{Bias of `uncorrected' temperatures due to spectral fitting as a function of $M_{500}$ at $z=0$. The level (and even the sign) of the bias depend on the details of the implemented sub-grid physics.}
\label{fig:Tbias}
\end{center}
\end{figure}

In Fig.~\ref{fig:hsebias}, we plot the median hydrostatic bias--$M_{500}$ relation for the various simulations, where the hydrostatic bias is defined as $\frac{M_{500,hse,spec}-M_{500}}{M_{500}}$. Consistent with previous simulation studies \citep[e.g.][]{Mathiesen1999,Rasia2006,Nagai2007b,Kay2012,Nelson2014}, we find a mean bias of $\sim- 20$ per cent for both groups and clusters. The scatter, which for clarity's sake is only shown for the \agn~8.0 model, increases with decreasing total mass.

In Fig.~\ref{fig:Tbias}, we plot the median bias of `uncorrected' temperatures due to spectral fitting as a function of $M_{500}$ for the various simulations. Previous studies \citep[e.g.][]{Mathiesen2001,Mazzotta2004,Rasia2006,Khedekar2013} found that the spectral temperatures are generally biased low compared to the mass-weighted temperatures. We find that the level (and even the sign) of the bias are dependent on the details of the sub-grid physics implementation, but defer a detailed analysis of the origin of hydrostatic and spectroscopic temperature biases to a future study (Le Brun et al.\ in preparation).

\bsp

\label{lastpage}

\end{document}